\documentclass[a4paper,11pt]{article}
\pdfoutput=1
\usepackage{jheppub}
\usepackage[usenames,dvipsnames]{xcolor}
\usepackage{slashed}
\usepackage[T1]{fontenc}
\usepackage[ansinew]{inputenc}
\usepackage{amsmath}
\usepackage{amssymb}
\usepackage{graphicx}
\usepackage{color}
\definecolor{darkblue}{cmyk}{0.9,0.9,0,0}
\definecolor{darkgreen}{rgb}{0,0.55,0}
\usepackage{minitoc}
\usepackage[normalem]{ulem}
\usepackage{enumitem}
\usepackage{tikz-cd}

\makeatletter

\newcommand{\comment}[1]{}

\newcommand{\beq}{\begin{equation}}
\newcommand{\eeq}{\end{equation}}
\newcommand{\beqq}{\begin{equation*}}
\newcommand{\eeqq}{\end{equation*}}
\newcommand\beqa{\begin{eqnarray}}
\newcommand\eeqa{\end{eqnarray}}
\newcommand\beqaa{\begin{eqnarray*}}
	\newcommand\eeqaa{\end{eqnarray*}}
\newcommand\bea{\begin{array}}
	\newcommand\eea{\end{array}}

\def\XXint#1#2#3{{\setbox0=\hbox{$#1{#2#3}{\int}$ }
		\vcenter{\hbox{$#2#3$ }}\kern-.5\wd0}}

\def\XXint#1#2#3{{\setbox0=\hbox{$#1{#2#3}{\int}$}
		\vcenter{\hbox{$#2#3$}}\kern-.5\wd0}}

\newcommand{\nn}{\nonumber}

\newcommand{\neqa}{\nonumber\end{eqnarray}}
\newcommand{\la}[1]{\label{#1}}

\newcommand{\hs}{\frac{\sqrt{3}}{2}}
\renewcommand{\d}{\partial}

\newcommand{\<}{{\langle}}
\renewcommand{\>}{{\rangle}}

\renewcommand{\v}{{\rm v}}

\newcommand{\re}{\relax{\rm I\kern-.18em R}}

\renewcommand{\sp}{p\hspace{-.40em}/}

\def\su2{{SU(2)}}

\def\[{\left[}
\def\]{\right]}

\def\s{\sigma}

\def\({\left(}
\def\){\right)}
\def\[{\left[}
\def\]{\right]}

\def\<{\langle}
\def\>{\rangle}

\def\cO{{\cal O}}

\def\cW{{\cal W}}
\def\cP{{\cal P}}
\def\cM{{\cal M}}

\def\s*{\ *_{\!\!\!\!\!\!\!\!\!\,_{\,_\text{\scriptsize{sym}}}}}
\def\hs*{\ \hat{*}_{\!\!\!\!\!\!\!\!\!\,_{\,_\text{\scriptsize{sym}}}}}
\def\d{\partial}

\def\i2{\frac{i}{2}}

\def\O{{\mathcal O}}

\def\spi{\relax{\rm \pi\kern-0.5em /}}
\def\sA{\relax{\rm A\kern-0.5em /}}
\def\sp{\relax{\rm p\kern-0.5em /}}
\def\sd{\relax{\rm \d\kern-0.5em /}}
\def\sk{\relax{\rm k\kern-0.5em /}}
\def\sn{\relax{\rm n\kern-0.5em /}}
\def\sl{\relax{\rm l\kern-0.5em /}}
\def\sP{\relax{\rm P\kern-0.7em /}}
\def\sBethe{\relax{\rm \Bethe\kern-0.5em /}}

\def\One{\mathbb{I}}

\newcommand{\Blue}[1]{{\color{blue}#1\color{black}}}

\def\be#1\ee{\begin{equation}\begin{aligned}
#1
\end{aligned}
\end{equation}}

\newcommand{\ii}{\mathrm{i}}
\newcommand{\dd}{\mathrm{d}}

\newcommand{\lo}{\overline{\mathcal{O}}}
\newcommand{\ro}{\mathcal{O}}

\newcommand{\llangle}{\<\!\<}
\newcommand{\rrangle}{\>\!\>}

\numberwithin{equation}{section}
\usepackage{varioref}
\usepackage{makeidx}
\usepackage{cleveref}
\makeindex

\title{\boldmath Bootstrapping Smooth Conformal Defects in Chern-Simons-Matter Theories}

\author[a]{Barak Gabai}
\author[b]{Amit Sever}
\author[c]{and De-liang Zhong}
\affiliation[a]{Laboratory for Theoretical Fundamental Physics, EPFL, Rte de la Sorge, CH-1015, Lausanne}
\affiliation[b]{School of Physics and Astronomy, Tel Aviv University, Ramat Aviv 69978, Israel}
\affiliation[c]{Theoretical Physics Group, The Blackett Laboratory, Imperial College London, Prince Consort Road London, SW7 2AZ, UK}

\abstract{The expectation value of a smooth conformal line defect in a CFT is a conformal invariant functional of its path in space-time. For example, in large $N$ holographic theories, these fundamental observables are dual to the open-string partition function in AdS. In this paper, we develop a bootstrap method for studying them and apply it to conformal line defects in Chern-Simons matter theories. In these cases, the line bootstrap is based on three minimal assumptions -- conformal invariance of the line defect, large $N$ factorization, and the spectrum of the two lowest-lying operators at the end of the line. On the basis of these assumptions, we solve the one-dimensional CFT on the line and systematically compute the defect expectation value in an expansion around the straight line. We find that the conformal symmetry of a straight defect is insufficient to fix the answer. Instead, imposing the conformal symmetry of the defect along an arbitrary curved line leads to a functional bootstrap constraint. The solution to this constraint is found to be unique.}

\begin{document} 
\maketitle
\flushbottom

\vspace{1.0cm}

\section{Introduction}

We study line defects in conformal field theory (CFT). Such one-dimensional operators undergo a defect Renormalization Group (RG) flow, \cite{Affleck:1991tk, Polchinski:2011im, Friedan:2003yc, Casini:2016fgb, Cuomo:2021rkm}. We focus on the fixed points of this flow, i.e., conformal line defects. Due to the absence of one-dimensional anomalies, such operators can be placed along an arbitrary smooth path without breaking their conformal symmetry. In particular, the expectation value of such an operator is a conformal invariant functional of the path. This functional is a fundamental probe of the theory. For example, it can serve as an order parameter for the confinement/deconfinement transition. In holographic large-$N$ gauge theories, the expectation value of a certain conformal defect is dual to the string partition function in AdS.

The only known examples where such a functional is explicitly known are in 2d YM \cite{Kazakov:1980zj}, and related to it, for a certain class of BPS lines on $S^2\subset{\mathbb R}^4$ in ${\cal N}=4$ SYM theory, \cite{Giombi:2009ms}. However, these examples are rather degenerate because the expectation value depends only on a two-dimensional area that is enclosed by the loop.\footnote{Stated differently, in these examples, the two-point function of the displacement operator and the cusp anomalous dimension are trivial.} See \cite{Drukker:1999zq, Polyakov:2000jg, Polyakov:2000ti, Semenoff:2004qr, Giombi:2009ek, Toledo:2014koa, Cooke:2017qgm} for more examples in which such functionals were studied at different limits. 

In this paper, we develop a bootstrap method to study smooth line defects. We apply it to one of the simplest, yet nontrivial, examples of conformal defects. These are the ones that are realized in the t' Hooft large $N$ limit of three-dimensional Chern-Simons (CS) theories with matter in the fundamental representation. In these cases, we formulate the bootstrap problem in terms of three assumptions. They consist of the conformality of the line defect, large $N$ factorization, and the spectrum of the two lowest-lying operators at the end of the line. We find that imposing the conformal symmetry of a flat (straight) defect is insufficient for fixing its expectation value when the line is not straight. On the other hand, imposing the conformal symmetry of the defect along an arbitrary smooth path leads to new nontrivial functional constraints.
In particular, they result in a unique bootstrap solution. We use it to solve the line CFT and compute the defect expectation value in a systematic expansion around the straight line. Our results are in perfect agreement with those that we have obtained in \cite{Gabai:2022vri,Gabai:2022mya} using an all-loop resummation of the perturbative expansion. Our consideration resembles those of conformal perturbation theory (see, e.g. \cite{Cardy:1996xt, Zamolodchikov:1987ti, Kutasov:1988xb, Ranganathan:1993vj, Gaberdiel:2008fn, Amoretti:2017aze, Behan:2017mwi, Cavaglia:2022yvv} for examples of the method and its many applications). For other methods of studying line defects that are based on conformal symmetry, see \cite{Hartman:2022zik,Poland:2022qrs} and references therein.

Famously, CS-matter theories possess a strong-weak bosonization duality. This duality maps the theory coupled to bosons to another theory coupled to fermions; see \cite{Gabai:2022mya} and the references therein. Our bootstrap axioms are blind to which of the two dual descriptions has been used to realize the line defects.  

The paper is organized as follows. In Section \ref{sec-setup} we set up the bootstrap problem. We detail the axioms on which it is based and our bootstrap strategy. As a warm-up, in Section \ref{sec-first order}, we compute the first-order variation of a mesonic line operator. This order follows directly from our axioms and does not require a bootstrap. We use it to set our notation and regularization scheme. In Section \ref{sec-second order} we bootstrap the second-order variation and use it to solve the line CFT. All the ingredients that appear at higher-order variations are present at the third order. In Section \ref{sec-third order} we bootstrap it and discuss its higher-order generalization. We end in Section \ref{sec-discussion} with a discussion.

\section{Setup}\la{sec-setup}

Our main objective in this paper is to determine the expectation value of mesonic line operators
\beq \label{mesonicline}
M[x(\cdot)]=\<\cM[x(\cdot)]\>=\<\lo(x_1){\cal W}[x(\cdot)]\ro(x_0)\>\,,
\eeq
where $\cal W$ is a conformal line operator along a smooth path, and $x(s)$ is some parameterization of the path. This path is oriented from the right boundary operator $\ro$ at $x_0\equiv x(0)$ to the conjugate left boundary operator $\lo$ at $x_1\equiv x(1)$. Instead of the mesonic line, we can consider the expectation value of a closed loop. As will become clear in the following, the latter is determined from the former.

Such operators are realized in CS theory with matter in the fundamental representation. For example, in CS theory with one fermion, $\cW$ can be a Wilson line in the fundamental representation that is stretched between fundamental and anti-fundamental fermion fields, \cite{Gabai:2022mya}.

We now detail three minimal properties of this observable, which we assume to hold throughout this paper. They were derived in \cite{Gabai:2022mya} by an explicit resummation of the 't Hooft perturbative expansion.

\subsection{Bootstrap Axioms}

\paragraph{Axiom I - Conformal invariance}
\hypertarget{axiom1}{}

The first axiom concerns the conformal invariance of the line operator $\cal W$. Under a conformal transformation, $x\to\tilde x$, the line operator transforms as
\beq \label{axiom1Eqn}
{\cal W}[x(\cdot)]\quad \mapsto \quad\widetilde{\cal W}[\tilde x(\cdot)]={\cal W}[\tilde x(\cdot)]\,.
\eeq

It follows from this axiom that an infinite straight line preserves an $SL(2,{\mathbb R})\times U(1)$ subgroup of the three-dimensional conformal group. Here, $SL(2,{\mathbb R})$ is the conformal symmetry of the line, and the $U(1)$ factor stands for rotations in the transverse plane to the line. Consequently, the operators on the line and those on which the line can end are classified by their (unitary) $SL(2,{\mathbb R})$ representation and their $U(1)$ transverse spin. Such operators can be divided into $SL(2,{\mathbb R})$ primaries and descendants. The latter can be obtained from the former using longitudinal derivatives. The primaries are characterized by their dimension $\Delta$, and their transverse spin $\mathfrak{s}$. We define the sign of the transverse spin to be correlated with the orientation of the line so that the spin of $x^+=(x^1+\ii x^2)/\sqrt 2$ is $+1$ for a line oriented in the $\hat x^3$ direction.

When the line is smooth (but not necessarily straight), we can zoom in close to a point on the line. Locally, it is approximately straight, and we can use the same classification of line and boundary operators as those on, and at the ends of, a straight line. 
\footnote{Namely, $\Delta$ and $\mathfrak s$ dictate how the operator transforms under the $SL(2,\mathbb{R})\times U(1)$ transformations that preserve the straight line that is tangent to the line at that point.} 

Operators with non-zero transverse spin are equipped with a transverse polarization vector $n$, with $\cO=\cO_{\Delta,{\mathfrak s}}(x,n)$. A transverse rotation of this polarization vector results in an overall phase factor of $e^{\pm\ii {\mathfrak s}\theta}$. The transverse spin of these operators may be factional. As a result, the line expectation value acquires a phase under $2\pi$ rotation of one of the boundary polarization vectors in the plane transverse to the line. In this paper, we will be concerned with the deformation of a straight line with an arbitrary normalization, which will be factored out. Therefore, keeping track of this $e^{\pm2\pi\ii{\mathfrak s}}$ phase is not relevant to us.\footnote{In order to have a single-valued expectation value, one can introduce a framing vector, which is a transverse polarization along the line. The boundary value of this vector is the boundary polarization vector, whereas the dependence of the mesonic line on the bulk value of this vector is topological. Having such a topological framing vector is natural from the perturbative definition of Wilson lines in CS theory \cite{Witten:1988hf}.}

\paragraph{Axiom II- Factorization}
\hypertarget{axiom2}{}

We are concerned with a ``large $N$ CFT'', where correlation functions factorize as 
\beq \label{factorizeexp}
\<\cM[x(\cdot),n]\,\widetilde \cM[\tilde x(\cdot),\tilde n]\>
=\<\cM[x(\cdot),n]\>\,\times\,\<\widetilde\cM[\tilde x(\cdot),\tilde n]\>\,\times\,\(1+\O(1/N)\)\,.
\eeq

Another large $N$ simplification is the factorization of line operators into a product of two boundary operators,
\beq\la{factorize}
\cO_\text{line}=\ro\times\lo\,.
\eeq
In large $N$ CS-matter theory, the operators $\cO_\text{line}$ transform in the adjoint representation, and the two boundary operators, $\ro$ and $\lo$, transform in the fundamental and anti-fundamental, respectively.\footnote{The singlet line operators are suppressed at large $N$.} Here, the transverse polarization vectors of the right and left boundary operators are the same.

Therefore, the properties of the line operators follow those of the left and right boundary operators in the product (\ref{factorize}). In particular, it follows from (\ref{factorizeexp}) and (\ref{factorize}) that the correlation function of operators on the line factorizes into a product of expectation values of mesonic line operators (\ref{mesonicline})
\beq
\<\ldots{\cal W}\cO_\text{line}{\cal W}\ldots\>=\<\ldots{\cal W}\ro\>\times \<\lo{\cal W}\ldots\>\,\times\,\(1+\O(1/N)\)\,.
\eeq

Note that axioms I and II do not imply that the correlation functions of operators on a straight line are trivial. Although their dependence on the positions of the operators factorizes, their spectrum and (properly normalized) overall coefficient are nontrivial. As a result, the dependence of $M$ on the shape of the path is nontrivial. We will bootstrap them using one more assumption about the low-lying operator spectrum.

\paragraph{Axiom III - Boundary spectrum} \hypertarget{axiom3}{}

The spectrum of boundary operators on the left end of the line is the same as that on the right, with the only difference being that the transverse spin is flipped,
\beq\la{spectrumL}
(\Delta_{\overline{\cO}},\mathfrak{s}_{\overline{\cO}})=(\Delta_{{\cO}},-\mathfrak{s}_{{\cO}})\,.
\eeq
This relation is inherited from the CPT symmetry of the theory, which simultaneously reflects the transverse plane and interchanges the left and right ends of the oriented line.

The third axiom concerns the spectrum of the low-lying $SL(2,{\mathbb R})$ primary boundary operators at the right end of the line. 
It states that the primary operator with $\mathfrak{s}\le{1\over2}$ and minimal conformal dimension is
\beq\la{firstop}
(\Delta_\cO,\mathfrak{s}_\cO)=(\Delta,1/2-\Delta)\,,
\eeq
and that the one with $\mathfrak{s}\ge{1\over2}$ is
\beq\la{secondop}
(\Delta_\cO,\mathfrak{s}_\cO)=(2-\Delta,3/2-\Delta)\,,
\eeq
where $\Delta\in[0,1]$. We also assume that the transverse spin of these primary operators is not degenerate. That is, all other primary operators have a different transverse spin than those in (\ref{firstop}) and (\ref{secondop}). This assumption can be traced back to having a CS-matter theory with a single flavor.

\subsection{Discussion} \label{sec:discussionSpinSource}

One direct consequence of the axioms is that the primary operator on the line with the lowest dimension is given by a factorized product of the right operator (\ref{firstop}) and its CPT conjugate
\beq
\cO_\text{bi-scalar}\equiv
\ro_{\Delta,1/2-\Delta}\times\lo_{\Delta,\Delta-1/2}\,,\qquad\text{with}\qquad\Delta_\text{bi-scalar}=2\Delta\,,\quad\mathfrak{s}_\text{bi-scalar}=0\,.
\eeq
For $\Delta\in[0,1/2)$, it is relevant, and therefore the conformal fixed point on the line is unstable. On the other hand, for $\Delta\in(1/2,1]$, that scalar primary operator is irrelevant and the fixed point is stable.\footnote{In the case where $\Delta=1/2$, the two fixed points collide. At that point, one sign of the coupling in front of $\cO_\text{bi-scalar}$ is marginally irrelevant and the other is marginally relevant.}

Chern-Simons theory with bosonic or fermionic fundamental matter has two different conformal line operators, one unstable and the other stable, \cite{Gabai:2022vri, Gabai:2022mya}. 
By allowing $\Delta$ to take a value in $[0,1]$, we cover both types of lines. In particular, the case where $\Delta=1/2$ corresponds to the free bosonic description, and the cases where $\Delta=0$ or $\Delta=1$ correspond to the free fermionic description. In these cases, the bootstrap deserves special care because, at these values, certain relevant or irrelevant operators become marginal. In this paper, we will avoid them by further assuming that $\Delta\ne0,1/2,1$. These values can still be obtained by analytic continuation. 

The axiom \hyperlink{axiom1}{I} \eqref{axiom1Eqn} states that the line is conformally invariant. Hence, the only source in a conformal transformation of the mesonic line operators comes from the boundary. These are taken to be $SL(2,{\mathbb R})$ primary operators of dimension $\Delta_\cO$ and transverse spin $\mathfrak{s}_\cO$. Consequently, if we transform both the path and the boundary polarizations, the only source comes from the boundary conformal weight
\beq\la{source}
{\widetilde\lo_{\Delta',\mathfrak{s}'}(\tilde x_1,\tilde n_1){\cal W}[\tilde x(\cdot)]\widetilde\ro_{\Delta,\mathfrak{s}}(\tilde x_0,\tilde n_0)\over\lo_{\Delta',\mathfrak{s}'}(x_1,n_1){\cal W}[x(\cdot)]\ro_{\Delta,\mathfrak{s}}(x_0,n_0)}=[\text{conformal factor}]\,.
\eeq

To compute this factor, we decompose the conformal transformation at the boundary of the line, $x\to\tilde x$, into translation, three-dimensional rotation, and dilation
\beq
\frac{\partial \tilde{x}^\mu}{\partial x^\nu} = \Omega(x) {\Lambda^\mu}_\nu(x)\,, \qquad \text{where} \quad \det {\Lambda^\mu}_\nu = 1\, .
\eeq
The dilatation $\Omega$ is part of the $SL(2,{\mathbb R})$ subgroup, under which the boundary operator has dimension $\Delta_\cO$. Hence,
\beq\la{conffactor}
[\text{conformal factor}]=\Omega(x_0)^{-\Delta}\times\Omega(x_1)^{-\Delta'}\,.
\eeq

One can also rewrite this equation in a form where the polarization vector is not transformed and the boundary spin dependence is manifest. Note that the rotation matrix $\Lambda\in SO(3)$ is, however, not a transverse rotation. Hence, the transverse planes before and after the conformal transformation are not the same. The only way that the boundary polarization vector on the transformed path would point in the same direction as that on the original path is if it points in the direction of the intersection between these two transverse planes. This is equivalent to starting with (\ref{source}) and rotating both $n$ and $\tilde n$, each in its transverse plane, so that they point in the same direction. Doing so results in the phase factor
\beq\la{spinfactor}
[\text{spin factor}]=e^{\ii{\mathfrak s}\theta_0}\times e^{\ii{\mathfrak s}'\theta_1}\,,
\eeq
where $\theta_0$ and $\theta_1$ are the total transverse rotation angles at the two end points. Let us focus on the right angle, $\theta_0$. We denote by $e_0$ and $\tilde e_0^\mu={\Lambda^\mu}_\nu e_0^\nu$ the unit tangent vectors before and after the conformal transformation. The deformation angle, $\theta_0$, is equal to the difference between two angles. One is the angle between $\tilde n$ and the intersection of the two transverse planes, $n^\text{int}\propto e_0\times\tilde e_0$, which comes from the numerator in (\ref{source}). From it we subtract the angle between $n$ and $n^\text{int}$, which comes from the denominator in (\ref{source}). This difference is defined in such a way that it vanishes when the conformal transformation is trivial. All together, we have
\beq\la{source2}
{\widetilde\lo_{\Delta',\mathfrak{s}'}(\tilde x_1,n_1^\text{int}){\cal W}[\tilde x(\cdot)]\widetilde\ro_{\Delta,\mathfrak{s}}(\tilde x_0,n_0^\text{int})\over\lo_{\Delta',\mathfrak{s}'}(x_1,n_1^\text{int}){\cal W}[x(\cdot)]\ro_{\Delta,\mathfrak{s}}(x_0,n_0^\text{int})}=[\text{conformal factor}]\times[\text{spin factor}]\,.
\eeq

Consider, for example, the case where the two boundary operators are primaries with the same conformal dimension and opposite transverse spin. In this case, the expectation value of the mesonic line operator takes the form 
\beq\la{generalform}
\<\lo_{\Delta,-\mathfrak{s}}(x_1,n_1){\cal W}[x(\cdot)]\ro_{\Delta,\mathfrak{s}}(x_0,n_0)\>={(2n_0^+n_1^-)^{\mathfrak{s}}\over|x_0-x_1|^{2\Delta}}\times F_{\Delta}[x(\cdot)]\,,
\eeq
where $F$ is a conformal invariant functional of the path that we would like to compute. Here, we have used light-cone coordinates to parameterize the transverse planes at the two endpoints, $\dd^2 s_\perp=2\dd x^+\dd x^-$.

\subsection{Notations}
We adopt the notation where the right and left boundary operators are denoted by
\beq\la{opnotation}
\cO^{(m,\ell)}\quad\text{and}\quad\overline{\cO}^{(m,\ell)}\,,
\eeq
where, $\ell$ is the transverse spin at $\Delta=1/2$. That is, in the free bosonic theory. Here, $m\ge0$ is the integer $SL(2,{\mathbb R})$ descendant level. The descendant operators have their dimension shifted by $m$ while their transverse spin remains the same as that of the primary. 

In this notation, the only primary operators that we assume to exist are
\beq\la{bottom}
\{\ro^{(0,0)},\ro^{(0,1)}\}\qquad\text{and}\qquad\{\lo^{(0,0)},\lo^{(0,-1)}\}\,.
\eeq 
Nevertheless, we adopt the general notation \eqref{opnotation}, informed by the retrospective understanding that the bootstrap constraints examined in this paper are consistent only if the spectrum of boundary operators comprises two non-degenerate, infinite towers of operators, characterized by increasing and decreasing transverse spins, respectively
\beq\la{spectrumR}
(\Delta^{(0,\ell)},\mathfrak{s}^{(0,\ell)})=\left\{\begin{array}{llcl}
\(\Delta,\frac{1}{2}-\Delta\)&\color{blue}+(|\ell|,\ell)&\quad&\color{blue}\ell\le0\, ,\\
\(2-\Delta,\frac{3}{2}-\Delta\)&\color{blue}+\(\ell-1,\ell-1\)&\quad&\color{blue}\ell\ge1\, ,
\end{array}\right.\,
\eeq
where \Blue{$\ell\in{\mathbb Z}$}. 

The spectrum given above is invariant under the transformation of the parameter $\Delta\to(2-\Delta)$ together with a sign flip of the transverse spin. This symmetry transformation interchanges the two towers. We denote it {\it tower-swap}.

We will also use the shorthand notation for the expectation value of mesonic lines with primary boundary operators
\beq\la{Mssbeq}
M^{(\ell,\ell')}_{s t}[x(\cdot)]\equiv\<\lo^{(0,\ell)}(x_s){\cal W}[x(\cdot)]\ro^{(0,\ell')}(x_t)\>\,,
\eeq
where we have dropped the dependence of the boundary operators on their transverse polarization vectors. In the case when the line is straight we will also drop the dependence of $M$ on $x(\cdot)$ and express (\ref{Mssbeq}) as
\beq
M^{(\ell,\ell')}_{s t}[x_\text{straight}(\cdot)]\equiv M^{(\ell,\ell')}_{s t}\equiv\<\!\<\lo^{(0,\ell)}(x_s)\,\ro^{(0,\ell')}(x_t)\>\!\>\,.
\eeq

\subsection{Bootstrap Strategy}

Using axioms \hyperlink{axiom1}{I}, \hyperlink{axiom2}{II} and \hyperlink{axiom3}{III}, we can now systematically bootstrap the expectation value of mesonic line operators along a smooth path. Our strategy is to start from a straight mesonic line, whose only nonzero expectation values are those that preserve the boundary spin
$\ell'=-\ell$. Their dependence on the length of the line is fixed by symmetry to be given by
\beq \label{Mssb}
M^{(\ell,\ell')}_{10}
= \delta_{\ell',-\ell}\, \frac{c_\ell}{|x_{10}|^{2\Delta_\ell}}\,,
\eeq
where $\Delta_\ell = \Delta(\ro^{(0,\ell)})$ is the dimension of $\ro^{(0,\ell)}$ and $c_\ell$ is a number that depends on our normalization of the boundary operators. It is convenient to choose the two boundary polarization vectors to be the same, so that the factor $2n_0^+n_1^-=n^2=1$ in (\ref{generalform}) and $c_l$ is real. 

We then smoothly deform the straight line
\beq\la{deformation}
x_\text{straight}(s)\quad\rightarrow\quad x(s)=x_\text{straight}(s)+\v(s)\,,
\eeq
and bootstrap $M[x(\cdot)]$ order by order in the smooth deformation profile $\v(s)/|x_1-x_0|$. More concretely, we express the deformed operator in terms of operator insertions along the straight line and bootstrap their coefficients. This allows us to explicitly compute the mesonic line expectation value, order-by-order, in the deformation.

To impose conformal symmetry on the expansion, we construct a one-parameter family of transformations, $x_s\to\tilde x_{\beta}(x_s)$. They can be any combination of translations, rotations, dilatation, and special conformal transformations, that are smoothly connected to the identity at $\beta=0$.\footnote{For example, a special conformal transformation can be parametrized as $\tilde x^\mu_{\beta}(x)=\frac{x^\mu - \beta^2\hat b^\mu (\hat b\cdot x)}{1-2\beta(\hat b\cdot x) +\beta^2\hat b^2 x^2}$, where $\hat b$ is some unit vector.} For convenience, we tune the deformation so that it deforms the endpoints in the transverse plane only. We then expand the deformed path $\tilde x_{\beta}(x_s)=x_s+\v_s(\beta)$ in powers of $\beta$ and reparameterize it so that at any point along the line the deformation vector is transverse to the original straight line, $\v_s(\beta)\cdot\dot x_s=0$. Because $\v_s(\beta)$ depends nonlinearly on $\beta$, this expansion in powers of $\beta$ is not the same as the expansions in powers of $\v(\beta)$. The two expansions, however, only start to differ at the third order (see Appendix \ref{sec:apd:confTransf} for details).

\section{First Order}\la{sec-first order}

Consider the variation of the path (\ref{deformation}) to first order in $\v/|x_1-x_0|$. It can be expressed as a linear combination of local operators inserted on the line and at the two boundaries
\beq \label{eqn-DeltaM1st}
\delta M=\lo\delta{\cal W}\ro+\delta \lo{\cal W}\ro+\lo{\cal W}\delta\ro\,.
\eeq
These boundary and line variations consist of all operators that are allowed by symmetry.

\subsection{Line Variation} 

The first order line variation can be brought to the form\footnote{In three dimension one can also write the structure $\epsilon_{\mu\nu\rho}\dot x^\mu \v^\nu \hat{\mathbb D}^\rho$. However, keeping in mind that the line is oriented, it is not independent of ${\mathbb D}$ in (\ref{displacement}). Explicitly, they are related by the redefinition ${\mathbb D}_\mu=\epsilon_{\mu\nu\rho}e^\nu\hat{\mathbb D}^\rho$, where $e^\mu=\dot x^\mu/|\dot x|$.}
\beq \label{displacement}
\delta W=\int\limits_0^1\dd s\,|\dot x_s|\v_s^\mu\, \cP[{\mathbb D}_\mu(x_s)\cW]\,,
\eeq
where we have used integration by parts to remove any derivatives from $\v$. The line operator ${\mathbb D}$ is called the \textit{displacement operator}. It follows from (\ref{displacement}) and axiom \hyperlink{axiom1}{I} that it is a dimension two primary operator with transverse spin equal to one. The form of this operator is uniquely fixed by the spectrum of lowest dimension operators, (\ref{firstop}), and (\ref{spectrumL}), to be given by
\beq \label{Dpm}
\begin{aligned}
{\mathbb D}_+(x)&=\eta\[\ro^{(0,1)}(x,n)\times\lo^{(0,0)}(x,n)\](\sqrt 2n_+)\,,\\
{\mathbb D}_-(x)&=\eta\[\ro^{(0,0)}(x,n)\times\lo^{(0,-1)}(x,n)\](\sqrt 2n_-)\,.
\end{aligned}
\eeq
Here, $n$ is an arbitrary unit transverse polarization vector and therefore $(\sqrt 2n_\pm)$ is a phase. The proportionality factor, $\eta$, depends on our choice of normalization of the boundary operators in (\ref{firstop}), (\ref{secondop}), and can conveniently be set to one. Here, we have chosen to keep it undetermined and identical for the two transverse directions.

Using (\ref{Mssb}) and the form of the displacement operator (\ref{Dpm}) we have that
\begin{align}
\<\!\<\ldots\,{\mathbb D}_-\,\ro^{(0,\ell)}\>\!\>&\propto\llangle\lo^{(0,-1)}\,\ro^{(0,\ell)}\rrangle=0 \qquad\text{for}\qquad\ell\ne1\,,\la{dmr}\\
\<\!\<\ldots\,{\mathbb D}_+\,\ro^{(0,\ell)}\>\!\>&\propto\llangle\lo^{(0,0)}\,\ro^{(0,\ell)}\rrangle\ =\ 0\qquad\text{for}\qquad \ell\ne0\,,\la{dpr}
\end{align}
where the dots stand for some left operator. Similarly, 
\begin{align}
\<\!\<\overline\cO^{(0,\ell)}\,{\mathbb D}_-\,\ldots\>\!\>&=0 \qquad\text{for}\qquad \ell\ne0\,,\la{dml}\\
\<\!\<\overline\cO^{(0,\ell)}\,{\mathbb D}_+\,\ldots\>\!\>&=0\qquad\text{for}\qquad\ell\ne-1\,.\la{dpl}
\end{align}

\subsection{Boundary Variation}
\label{sec:bdrVar}

The boundary variation may include derivatives of $\v$,
\beq \label{boundartvar}
\delta \cO=\v^\mu\delta_{\mu}^{(0)}\cO+\dd \v^\mu\delta_{\mu}^{(1)}\cO+\dd \dd \v^\mu\delta_{\mu}^{(2)}\cO+\dots\,.
\eeq
Here, $\dd\v=\dot\v/|\dot x|$, is the {\it path derivative} of the variation vector $\v$ at the boundary. Similarly, $\dd \dd\v=\ddot\v/|\dot x|^2-\dot\v\d_s|\dot x|/|\dot x|^3$, etc., are the higher-order reparametrization invariant derivatives of $\v$ along the path. All terms on the right-hand side of (\ref{boundartvar}) must have the same conformal dimension and spin as those of $\cO$. Because $\Delta(\v)=-1$, $\Delta(\dd\v)=0$, $\Delta(\dd \dd\v)=1$, etc., it follows that the dimension of the boundary operators $\delta_\mu^{(n)}\cO$ grows with $n$. For a transverse variation, the operators $\delta_\mu^{(n)}\cO$ have one more unit of transverse spin with respect to that of $\cO$, while for a longitudinal variation, it remains the same. Note that because we have already fixed the line variation to be given by (\ref{displacement}), we no longer have the freedom to shift terms between $\delta\cO$, $\delta\overline\cO$, and $\delta\cW$ in (\ref{boundartvar}) using integration by parts.

We refer to the operator $\delta_{\mu}\cO\equiv \delta_{\mu}^{(0)}\cO$ as the {\it path derivative} of the boundary operator $\cO$. As we go to higher orders in the path deformation, we will have more boundary and line operator insertions. We denote the boundary operator with higher powers of the deformation as
\beq\la{multideformation}
\v^{\mu_1}\v^{\mu_2}\dots \v^{\mu_m}\delta_{\mu_1}\delta_{\mu_2}\dots\delta_{\mu_m}\cO\,.
\eeq
Importantly, this multi-path derivative is not a local deformation of the path. Instead, it is the operator that stands in front of $\v^{\mu_1}\v^{\mu_2}\dots \v^{\mu_m}$ in a \textit{smooth} variation of the path. Hence, unlike covariant derivatives, these path variations commute by definition,
\beq
\delta_\mu\delta_\nu\cO=\delta_\nu\delta_\mu\cO\ne\delta_\nu\(\delta_\mu\cO\)\,.
\eeq

Using path derivatives, the descendant boundary operators are defined as
\beq\la{descendant}
\ro^{(n+1,\ell)}=\delta_3\ro^{(n,\ell)}\,,\qquad \lo^{(n+1,\ell)}=\delta_3\lo^{(n,\ell)}
\eeq
where $\delta_3$ is the longitudinal path derivative. It is oriented in the direction of the path, from $\ro$ to $\lo$.

\subsection{Boundary Counter-terms}
\label{sec:KeyHole}

The perturbative expansion of the mesonic line operator around the straight line (\ref{deformation}) is a conformal perturbation theory, parameterized by the deformation profile. To carry it out order by order in $\v(s)$ we have to introduce a UV cutoff and renormalize the line. We denote by $\epsilon$ the corresponding UV cutoff, which has the dimension of length. 

The resulting structure of the boundary variations is largely constrained by symmetry. Consider, for example, the operator $\delta_-\ro^{(0,1)}$. It has transverse spin $\mathfrak{s} = \mathfrak{s}(\ro^{(0,1)})-1=1/2-\Delta$. According to axiom \hyperlink{axiom3}{III}, the only boundary operators with this transverse spin are $\ro^{(0,0)}$ and its descendants $\ro^{(n,0)}$. However, their conformal dimension differs from that of $\delta_-\ro^{(0,1)}$ by a non-integer value $\Delta(\ro^{(n,0)})-\Delta(\delta_-\ro^{(0,1)})=n-3+2\Delta$. Therefore, the decomposition of $\delta_{-}\ro^{(0,1)}$ in terms of boundary operators takes the form
\beq \label{pathvar01}
\delta_{-}\ro^{(0,1)}=\frac{1}{\epsilon^{3-2\Delta}}\sum_{n=0}^\infty 
b_n\epsilon^n\,\ro^{(n,0)}\,,
\eeq
where all coefficients, $\{b_n\}_{n=0}^\infty$, are scheme dependent. 

The divergent terms in (\ref{pathvar01}) are counter-terms. Their coefficients are fixed so that they cancel out divergences that result from the integration of the displacement operator. For the case of $\delta_{-}\ro^{(0,1)}$, this integral takes the form
\beq
\<\!\<\ldots\,\delta\cW\,\ro^{(0,1)}\>\!\>=
\int\limits_0\dd s\,|\dot x|\,\<\!\<\ldots\,\[\v_s^+{\mathbb D}_+(x_s)+\v_s^-{\mathbb D}_-(x_s)\]\,\ro^{(0,1)}(x_0)\>\!\>\,,
\eeq
where the dots stand for some left boundary operator. Only the term with ${\mathbb D}_-$ contributes to the expectation value; see (\ref{dmr}). As a result, the integral may have a local divergence coming from the $s\to0$ region of integration. 

We choose a proper time parameterization of the straight line 
\beq\la{straightparametrization}
x_\text{straight}(s)=x_0+s(x_1-x_0)\,, 
\eeq
and use a point splitting regulator, $\ell>\tilde\epsilon=|x_{10}|\epsilon$. With this choice, the divergence takes the form
\begin{align}
&|x_{10}|\int\limits_{\tilde\epsilon}\dd s\,\v_s^+\,\<\!\<\ldots\,{\mathbb D}_-(x_s)\,\ro^{(0,1)}(x_0)\>\!\>\nn\\
=&\eta|x_{10}|\int\limits_{\tilde\epsilon}\dd s\,\v_s^+\,\<\!\<\ldots\,\ro^{(0,0)}(x_s)\>\!\>M_{s0}^{(-1,1)}\\
=&{\eta\,c_{    -1}\over|x_{10}|^{3-2\Delta}}\left[\v_0^+\<\!\<\ldots\,\ro^{(0,0)}(x_0)\>\!\>\int\limits_{\tilde\epsilon}{\dd s\over s^{4-2\Delta}}\right.\nn\\
&\qquad\qquad\ \ \left.+\(\dot \v_0^+\<\!\<\ldots\,\ro^{(0,0)}(x_0)\>\!\>+\v_0^+|x_{10}|\<\!\<\ldots\,\ro^{(1,0)}(x_0)\>\!\>\)\int\limits_{\tilde\epsilon}{\dd s\over s^{3-2\Delta}}+\ldots\right]\,.\nn
\end{align}

The divergences $\v_0^+/\epsilon^{3-2\Delta}$ and $\v_0^+/\epsilon^{2-2\Delta}$ are canceled by setting $b_0=-\eta c_{-1}/(3-2\Delta)$ and $b_1=-\eta c_{-1}/(2-2\Delta)$ in (\ref{pathvar01}). Namely, we have
\beq
\label{eq:explicit1var}
\delta_-\ro^{(0,1)}=-{\eta c_{-1}\over(3-2\Delta)\epsilon^{3-2\Delta}}\ro^{(0,0)}-{\eta c_{-1}\over(2-2\Delta)\epsilon^{2-2\Delta}}\ro^{(1,0)}+O(\epsilon^{2\Delta-1})\,.
\eeq
Similarly, the $\dot \v_0^+/\epsilon^{2-2\Delta}$ divergences are canceled by setting in (\ref{boundartvar})
\beq
\delta_-^{(1)}\ro^{(0,1)}=-{\eta c_{-1}\over(2-2\Delta)\epsilon^{2-2\Delta}}\ro^{(0,0)}+O(\epsilon^{2\Delta-1})\,.
\eeq
In this way, all the divergent terms in $\delta_-\ro^{(0,1)}$ are fixed. 

Similarly, for $\delta_{+}\ro^{(0,0)}$ we have
\beq \label{pathvar00}
\delta_{+}\ro^{(0,0)}=\frac{1}{\epsilon^{2\Delta-1}}\sum_{n=0}^\infty 
\tilde b_n\epsilon^n\,\ro^{(n,1)}\,.
\eeq
The coefficients of the divergent terms are tuned to cancel those that result from the integration of ${\mathbb D}_+$ in $\<\!\<\ldots\,\delta\cW\,\ro^{(0,0)}\>\!\>$.

\subsection{The Keyhole Prescription}
\label{sec:keyhole}

Consider, for example, the first-order variation
\beq\la{M01var}
\delta M^{(0,1)}=\llangle\lo^{(0,0)}\delta\cW\ro^{(0,1)}\rrangle+\llangle\delta\lo^{(0,0)}\ro^{(0,1)}\rrangle+\llangle\lo^{(0,0)}\delta\ro^{(0,1)}\rrangle\,,
\eeq
In this case, the corresponding decomposition of $\delta\lo^{(0,0)}$ and $\delta\ro^{(0,1)}$ in terms of boundary operators (\ref{boundartvar}) comes with fractional powers of $\epsilon$. As discussed above, the role of these boundary operators is to remove divergences that result from the integration of the displacement operator near the boundary in
\beq\la{bulkint}
\<\!\<\lo^{(0,0)}\,\delta\cW\,\ro^{(0,1)}\>\!\>=|x_{10}|\int\limits_{\tilde\epsilon}^{1-\tilde\epsilon}\dd s\,\v_s^-\,\<\!\<\lo^{(0,0)}\,{\mathbb D}_-(x_s)\,\ro^{(0,1)}\>\!\>\,.
\eeq
This integral takes the form
\beq\la{cutint}
\<\!\<\lo^{(0,0)}\,\delta\cW\,\ro^{(0,1)}\>\!\>={\eta\, c_0\, c_{-1}\over|x_{10}|^3}\int\limits_{\tilde\epsilon}^{1-\tilde\epsilon}\dd s\,{\v_s^-\over(1-s)^4}\({1-s\over s}\)^{2\Delta}\,.
\eeq
It has a branch cut of order $\Delta$, with branch points $s=0$ and $s=1$. The result of regularizing this integral and canceling the divergences with the boundary variations in (\ref{M01var}) is equivalent to letting the cut run between the two branch points and converting this integral into a contour integral around it
\beq\label{keyhole}
\delta M^{(0,1)}_{10}={\eta |x_{10}|\over{2 \ii}\sin(2\pi\Delta)} \oint\limits_{[0,1]} \dd s\, \v_s^- 
M^{(0,0)}_{1s}M^{(-1,1)}_{s0}+[\text{positive powers of }\epsilon]\,,
\eeq
where we have assumed that the function $\v^-(s)$ does not have singularities along the segment $s\in[0,1]$. For convenience, we have also assumed that the variation at the boundary acts transversely to the line, $\v_0^3=\v_1^3=0$. Longitudinal boundary variations act trivially by rescaling the length of the line, $|x_{01}|$ in (\ref{Mssb}). The cancellation of divergent boundary terms against the terms in (\ref{pathvar00}) and the conjugate of (\ref{pathvar01}) can be understood as providing the cup or keyhole around the end of the cut, making the integral finite.

We define the keyhole regularization prescription as\footnote{For generic $\Delta$, all divergent terms near the boundary are power-like, therefore there is no ambiguity in the finite part in the $\epsilon \rightarrow 0$ limit, which only happens for a logarithmic divergence.}
\beq\label{keyhole2}
\delta M^{(0,1)}_{10}={\eta |x_{10}|\over{2 \ii }\sin(2\pi\Delta)} \oint\limits_{[0,1]} \dd s\, \v_s^- 
M^{(0,0)}_{1s}M^{(-1,1)}_{s0}={\eta\,c_0\, c_{-1} |x_{10}|\over{2 \ii }\sin(2\pi\Delta)} \oint\limits_{[0,1]} \dd s\,{\v_s^-\over(1-s)^4}\({s-1\over s}\)^{2\Delta}\,,
\eeq
without additional positive powers of $\epsilon$. This prescription amounts to a certain choice of coefficients in the expansion of the boundary variation. Apart from divergent counterterms, these series consist of boundary operators with increasing dimensions and (fractional) powers of $\epsilon$; see, for example, (\ref{pathvar01}). In the keyhole prescription, they are fixed so that (\ref{keyhole2}) holds at finite $\epsilon$.

At higher orders in the line deformation, these scheme-dependent terms with a positive power of $\epsilon$ may combine with divergent ones to give finite contributions. The total finite part, however, is scheme-independent. In Appendix \ref{apd:sec:2ndOrderSchemeInd}, we demonstrate this explicitly for two different schemes.

Similarly to (\ref{keyhole2}), the keyhole prescription for $\delta M^{(-1,0)}$ reads
\beq \label{keyhole3}
\delta M^{(-1,0)}_{10}={\ii \eta |x_{10}|\over2\sin(2\pi\Delta)} \oint\limits_{[0,1]} \dd s\, \v_s^+ M^{(-1,1)}_{1s} M^{(0,0)}_{s0}\,.
\eeq

In particular, for a polynomial $\v_s^-\propto s^n$ in (\ref{keyhole}), we end up with the integral
\beq \label{eqn-oneVar-KeyholeFinite}
\frac{1}{2\ii \sin 2\pi \Delta}\oint\limits_{[0,1]} \dd s \frac{s^{n}}{s^{2\Delta} (s-1)^{2(2-\Delta)}} = -\frac{\Gamma (2 \Delta -3) \Gamma (n-2 \Delta +1)}{\Gamma (n-2)} \, .
\eeq
Note the factor of $\Gamma(n-2)$ in the denominator. It means that $\delta M^{(0,1)}=0$ for $\v^-\propto s^0$, $\v^-\propto s^1$, and $\v^-\propto s^2$. These correspond in turn to global translation, rotation, and special conformal transformations, which are indeed symmetries of the mesonic line expectation value. To be precise, $\lo^{(0,0)}\cW\ro^{(0,1)}$ transforms covariantly under conformal transformations, with non-zero weight at the boundaries, (\ref{source}). These weights only start to contribute at second order and are analyzed in the next section. Hence, to the linear order, $M^{(-1,0)}$ is invariant under these transformations. There are no further constraints imposed by conformal symmetry at this order for mesonic line operators that end on the four operators in (\ref{bottom}).

We can now use (\ref{keyhole2}) and (\ref{keyhole3}) to compute the first-order variation of $M^{(0,1)}$ and $M^{(-1,0)}$ as functional of the shape of the path. The overall coefficient depends on our normalization of these operators. This freedom is reflected in the factor of $\eta$ that we have chosen to keep.

\section{Second Order}\la{sec-second order}

In the second order, for the first time, finite terms with undetermined coefficients enter the line variation. We list them explicitly in Section \ref{linevar2nd} and then bootstrap them in Section \ref{bootstrapsec2nd}.

\subsection{Line Variation}\la{linevar2nd}

The second-order variation can be divided into line and boundary contributions as 
\beq \label{eqn-Var-2ndOrd}
\begin{aligned}
\delta^2\cM & =\delta^2 \bar{O}\,\mathcal{W}\, O +\delta\,\bar{O}\,\mathcal{W}\,\delta O +\bar{O}\,\mathcal{W}\,\delta^2 O\, \\
& +\delta\bar{O}\,\delta\mathcal{W}\,O+\bar{O}\,\delta\mathcal{W}\,\delta O+\bar{O}\, \delta^2\mathcal{W}\,O\,.
\end{aligned}
\eeq

Because the dynamics on the line is that of a local CFT, any term in which the two variations act at two separated points on the line is given by two first-order variations. New operators appear when the two variations act at the same point, either at the boundary or in the bulk of the line. At the boundary of the line, the second-order variation takes a similar form to the first-order one in (\ref{boundartvar}) and is given by
\beq \label{boundartvar2}
\delta^2 \cO=\v^\mu \v^\nu\delta_{\mu}^{(0)}\delta_{\nu}^{(0)}\cO+\dd \v^\mu \v^\nu\delta_{\mu}^{(1)}\delta_{\nu}^{(0)}\cO+\dd \v^\mu \dd \v^\nu\delta_{\mu}^{(1)}\delta_{\nu}^{(1)}\cO+\dd \dd \v^\mu \v^\nu\delta_{\mu}^{(2)}\delta_{\nu}^{(0)}\cO+\dots\,,
\eeq
where in our parametrization (\ref{straightparametrization}), $\dd \v=\dot\v/|x_{10}|$, etc, see definition below \eqref{boundartvar}. At the bulk of the line, the second order line variation reads
\beq \label{eqn-Var-2nd-2Var}
\delta^2 \mathcal{W} =|x_{10}|^2 \int\limits_0^1\dd s\,\v_s^\mu\int\limits_0^1\dd t\,\v_t^\nu\,\cP[{\mathbb D}_\mu(x_s){\mathbb D}_\nu(x_t)\cW]+|x_{10}|\int\limits_0^1\dd s\,\cP[\delta{\mathbb D}(x_s)\cW]\,.
\eeq
Here, using integration by parts, we have chosen the double integral to run over two displacement operators, with no derivatives of $\v$. With this choice, the operator $\delta{\mathbb D}$ takes a form similar to $\delta^2 \cO$ in (\ref{boundartvar2}). That is, it is a sum of operators that are weighted by $\v^\mu_s \v^\nu_s$, and higher derivatives such as $\v^\mu_s \dd \v^\nu_s$, etc. 

We can now repeat the analogous treatment to the first-order variation, canceling divergences using counter-terms and keeping all terms that are finite at $\epsilon\to0$. The conclusion of this tedious computation is, however, very simple. We may list only the regularized integrals and boundary terms that are finite as $\epsilon\to0$ and can appear in (\ref{eqn-Var-2ndOrd}). The rest of the terms are either divergent and are tuned to cancel out or go to zero as the UV cutoff is removed. The coefficients in front of the finite terms are either fixed from the first-order variation or are new, and therefore left unfixed. In the next section, we will bootstrap them using the axioms.

In what follows, we will first focus on the case where $M=M^{(0,0)}$. As before, we do not consider longitudinal boundary variations because they act trivially.

\subsubsection{Boundary Terms with One Variation}

The boundary term with one variation on the right and the other on the left is given by \beq\la{BBvar}
\begin{aligned}
\llangle \delta\lo^{(0,0)}\delta\ro^{(0,0)}\rrangle&=\llangle(\v_1^-\delta_-+\dd \v_1^-\delta_-^{(1)}+\dots)\lo^{(0,0)}(\v_0^+\delta_++\dd \v_0^+\delta_+^{(1)}+\dots)\ro^{(0,0)}\rrangle\\
&+\llangle(\v_1^+\delta_++\dd \v_1^+\delta_+^{(1)}+\dots)\lo^{(0,0)}(\v_0^-\delta_-+\dd \v_0^-\delta_-^{(1)}+\dots)\ro^{(0,0)}\rrangle\,.
\end{aligned}
\eeq

The operators in the first line have expansions of the form (\ref{pathvar00}), with fractional powers of $\epsilon$. The ones with negative powers of $\epsilon$ are counter-terms that cancel divergences arising from the integrals considered below. In this order, those with positive powers of $\epsilon$ vanish as $\epsilon\to0$. In summary, none of the terms in the first line of (\ref{BBvar}) leads to a finite contribution. 

The terms in the second line are operators with transverse spins $\mathfrak{s}=-\bar{\mathfrak{s}}=(1/2-\Delta)-1$. Therefore, if they exist, they are new operators. According to the axiom \hyperlink{axiom3}{III}, their dimension must be strictly larger than that of $\cO^{(0,0)}$. In particular, the operators $\delta_-^{(l)}\ro^{(0,0)}$ and $\delta_+^{(l)}\lo^{(0,0)}$ with $l\ge1$ come with a positive power of $\epsilon$ and do not contribute as $\epsilon\to0$. 

We remained with the boundary operator 
\beq\la{dm00}
\delta_-\ro^{(0,0)}\equiv\epsilon^{\Delta^{(0,-1)}-\Delta^{(0,0)}-1}\Big(\ro^{(0,-1)}+\sum_{n=1}^\infty a_n\epsilon^n\ro^{(n,-1)}\Big)\,,
\eeq
and its CPT conjugate. The power of $\epsilon$ cannot be negative because there is no other contribution at that transverse spin that can cancel with it. So $\Delta^{(0,0)}<\Delta^{(0,-1)}\le\Delta^{(0,0)}+1$. The operator $\ro^{(0,-1)}$ appears multiplying $\v^-_0$ in (\ref{BBvar}) and is the operator of the lowest dimension with that transverse spin. Therefore, if it exists, it is an $SL(2,{\mathbb R})$ primary. The operators $\ro^{(n,-1)}$ are the corresponding $SL(2,{\mathbb R})$ descendants and the coefficients $a_n$ are scheme dependent. For the second line in (\ref{BBvar}) to yield a finite contribute, the operator $\ro^{(0,-1)}$ must have conformal dimension
\beq\la{Deltam1}
\Delta^{(0,-1)}=\Delta^{(0,0)}+1=\Delta+1\,.
\eeq
Below we will find that, indeed, such an operator must exist. Assuming that it does, equation (\ref{BBvar}) reduces to
\beq\la{BBvar2}
\llangle \delta\lo^{(0,0)}\delta\ro^{(0,0)}\rrangle=\v_1^+\v_0^-M_{10}^{(1,-1)} +[\text{positive powers of }\epsilon]+[\text{negative powers of }\epsilon]\,,
\eeq
where
\beq\la{c1definition}
M_{10}^{(1,-1)}\equiv\llangle \lo^{(0,1)}\ro^{(0,-1)}\rrangle\equiv {c_1\over|x_{10}|^{2\Delta+2}}\,.
\eeq
and $\lo^{(0,1)}$ is the CPT conjugate of $\ro^{(0,-1)}$.

\subsubsection{Boundary Terms with Two Variations}

Next, we have new boundary terms corresponding to second-order variation at the same boundary points, one in the plus direction and the other in the minus direction. These have the same transverse spin as that of $\ro^{(0,0)}$ and include all possible scalar operators $\ro^{(n,0)}$, weighted by the appropriate power of the cutoff. Because the spectrum at that transverse spin is non-degenerate, no other new operators can appear. We end up with
\begin{align}\label{eq:2ndOrderMarginal1}
\llangle \bar{O}^{(0,0)} \delta^2 O^{(0,0)} \rrangle= &\Big[ \v_0^+ \v_0^-\big({a_0\over\epsilon^2}+{a_1\over\epsilon}\delta_3+\gamma_0\delta_3^2\big)+\(\v_0^+ \dd \v_0^- + \v_0^- \dd \v_0^+\)\big({a_2\over\epsilon}+\gamma_1\delta_3\big)\\
&+\gamma_2 \(\v_0^+ \dd \dd \v_0^- + \v_0^- \dd \dd \v_0^+\)+\gamma_3 \dd \v_0^+ \dd \v_0^- \nn\\
&+\(\v_0^+ \dd \v_0^- - \v_0^- \dd \v_0^+\)\big({a_3\over\epsilon}+\gamma_4\delta_3\big)+\gamma_5 \(\v_0^+ \dd \dd \v_0^- - \v_0^- \dd \dd \v_0^+\)\Big]M_{10}^{(0,0)}\,,\nn
\end{align}
where $\delta_3$ is the longitudinal path derivative (\ref{descendant}). The coefficients $a_i$ are counter terms that are tuned to cancel the divergences of the integrals. 

Similarly, we have 
\begin{align}\label{eq:2ndOrderMarginal12}
\llangle \delta^2\bar{O}^{(0,0)}  O^{(0,0)} \rrangle= &\Big[ \v_1^+ \v_1^-\big({\tilde a_0\over\epsilon^2}-{\tilde a_1\over\epsilon}\delta_3+\tilde\gamma_0\bar \delta_3^2\big)+\(\v_1^+ \dd \v_1^- + \v_1^- \dd \v_1^+\)\big({\tilde a_2\over\epsilon}-\tilde\gamma_1\delta_3\big)\\
&+\tilde\gamma_2 \(\v_1^+ \dd \dd \v_1^- + \v_1^- \dd \dd \v_1^+\)+\tilde\gamma_3 \dd \v_0^+ \dd \v_0^- \nn\\
&-\(\v_1^+ \dd \v_1^- - \v_1^- \dd \v_1^+\)\big({\tilde a_3\over\epsilon} -\tilde\gamma_4\delta_3\big)-\tilde\gamma_5 \(\v_1^+ \dd \dd \v_1^- - \v_1^- \dd \dd \v_1^+\)\Big]M_{10}^{(0,0)}\,,\nn
\end{align}
where the tilted coefficients are independent of the un-tilted ones.\footnote{Recall that CPT symmetry exchanges the right and left ends of the line and the transverse spin. Therefore, as long as we choose a regularization scheme that is compatible with this symmetry, we are guaranteed to have $\tilde a_i=a_i$ and $\tilde\gamma_i=\gamma_i$.
In what follows, we will not work with a symmetric scheme and these coefficients will be different.}

\subsubsection{Single Integral}
\label{sec:2ndOrder:singleInt}

There are two types of terms with a single integral. One consists of one boundary and one bulk variation, given by $ \langle \delta \bar{O} \delta \mathcal{W} O \rangle + \langle \bar{O} \delta \mathcal{W} \delta O \rangle$ in (\ref{eqn-Var-2ndOrd}). The other type of terms with a single integral have two variations at the same point in the bulk, given by the last term in (\ref{eqn-Var-2nd-2Var}). We now consider these contributions in turn.

The first type does not lead to a finite contribution that depends on $\v_s$ at $1>s>0$. It does, however, lead to finite contributions concentrated at the boundaries that can be absorbed in a shift of the coefficients $\gamma_i$, $\tilde\gamma_i$ in (\ref{eq:2ndOrderMarginal1}), (\ref{eq:2ndOrderMarginal12}). 

Explicitly, the term $\langle \bar{O}^{(0,0)}\delta \mathcal{W} \delta O^{(0,0)} \rangle$ takes the form
\beq
\begin{aligned}\label{eq:onebulkr}
\langle \bar{O}^{(0,0)} \delta \mathcal{W} \delta O^{(0,0)} \rangle & = \eta \,|x_{10}|\int \dd s\ \llangle \bar{O}^{(0,0)}_1 O^{(0,0)}_s \rrangle\,  \v_s^-\, \llangle \bar{O}^{(0,-1)}_s \delta O^{(0,0)}_0 \rrangle \\
&= \frac{\eta\,|x_{10}|}{\epsilon^{2\Delta-1}}  \int \dd s\,M_{1s}^{(0,0)}\,  \v_s^-\, \llangle \bar{O}^{(0,-1)}_s  \big(\v_0^+ \tilde b_0 O^{(0,1)}_0 +\cO(\epsilon)\big)\rrangle\,,
\end{aligned}
\eeq
where in the first line we have used the form of the displacement operator \eqref{Dpm}, and in the second line we used the expansion of $\delta_+\ro^{(0,0)}_0$ in terms of the operators $\ro^{(n,1)}_0$, see (\ref{pathvar00}). Similarly, we have
\beq
\begin{aligned}\label{eq:onebulkl}
\langle \delta\bar{O}^{(0,0)} \delta \mathcal{W} O^{(0,0)} \rangle & = \eta\,|x_{10}| \int \dd s\ \llangle \delta\bar{O}^{(0,0)}_1 O^{(0,1)}_s \rrangle\,  \v_s^+\, \llangle \bar{O}^{(0,0)}_sO^{(0,0)}_0 \rrangle \\
&= \frac{\eta\,|x_{10}|}{\epsilon^{3-2\Delta}}  \int \dd s\,\llangle \big(\v_0^-\,b_0\bar{O}^{(0,-1)}_0  +\cO(\epsilon)\big) O^{(0,1)}_s \rrangle \v_s^+\,M_{s0}^{(0,0)}\,,
\end{aligned}
\eeq
where we used the expansion of the boundary operator (\ref{pathvar01}).

We see that these two integrands come with a fractional power of $\epsilon$. The coefficients of the divergent terms, $\eta b_0$ and $\eta\tilde b_0$, are tuned to cancel the divergences that come from the double integral that we consider in the next subsection. Hence, the integrals (\ref{eq:onebulkr}) and (\ref{eq:onebulkl}) do not lead to finite bulk contributions. 

As for the first-order variation in Section \ref{sec:KeyHole},
these integrals lead to boundary terms that can be divergent or finite. Divergences are canceled by the boundary counterterms in (\ref{BBvar2}) and \eqref{eq:2ndOrderMarginal1}. The finite boundary terms can be absorbed in a shift of the $\gamma_i$'s in (\ref{eq:2ndOrderMarginal1}).

The other type of one integral has two variations at the same point on the line and is given by 
\beq
\begin{aligned}
\llangle \lo^{(0,0)}\delta{\mathbb D}(x_s)\ro^{(0,0)}\rrangle&={1\over\epsilon^{3-2\Delta}}\llangle \lo^{(0,0)}\big[d_1\,\v_s^+\v_s^-\ro^{(0,0)}_s\times\lo^{(0,0)}_s+O(\epsilon)\big]\ro^{(0,0)}\rrangle\\
&+\Xi\(\dd \v_s^+\dd\dd \v_s^--\dd \v_s^-\dd\dd \v_s^+\)\llangle \lo^{(0,0)}\ro^{(0,0)}\rrangle\,,
\end{aligned}
\eeq
where $d_1$ and $\Xi$ are numerical coefficients. Similarly to (\ref{eq:onebulkl}), the integrand in the first line only has fractional powers of $\epsilon$ and, therefore, does not contribute a finite bulk contribution. The integrand in the second line is proportional to the identity operator on the line. It measures the local winding of the deformation and is scheme-independent. It can be changed using integration by parts and a shift in the boundary terms. Here we have chosen to write it in an anti-symmetric way between the plus and minus directions.\footnote{The sum $\dd \v_s^+ \dd \dd \v_s^- + \dd \v_s^- \dd \dd \v_s^+$ is a total derivative on the line and thus, can be absorbed into the boundary terms. Consequently, only the anti-symmetric combination would be the genuine operator on the line.}

\subsubsection{Double Integral}

The double integral takes the form 
\begin{align}\la{doubleint}
\delta^2 M_{10}^{(0,0)} \Big|_{\text{2-int}}^\text{bare}\equiv&\,\langle \bar{O}^{(0,0)} \delta^2 \mathcal{W}\, O^{(0,0)}\rangle-|x_{10}|\int\limits\dd s\, \llangle\lo^{(0,0)}\,\delta{\mathbb D}(x_s)\ro^{(0,0)}\,\rrangle\\
=&\,\eta^2\,|x_{10}|^2 \iint\limits_{s>t} \dd s\, \dd t\, \v_s^-\v_t^+ \, M_{1s}^{(0,0)}  M_{st}^{(-1,1)}  M_{t0}^{(0,0)}\, .\nn
\end{align}
We shall now choose a regularization scheme and tune the counter-terms in the one integral and boundary variation to arrive at a finite regularized double integral.

For fixed $s$, the $t$-integral is the one we have considered before at first order. We choose to regulate it using the keyhole prescription. We remain with the $s$-integral over $\v_s^-M_{1s}^{(0,0)}\delta M_{s0}^{(-1,0)}$, where $\delta M_{s0}^{(-1,0)}$ is the keyhole-regulated integral in (\ref{keyhole3}). The regions of integration near the boundaries may still lead to divergences. Naively, close to $s=0$ we have $\delta M_{s0}^{(-1,0)}\sim \v^+_0/s^3$. However, using the fact that the keyhole integral in (\ref{eqn-oneVar-KeyholeFinite}) vanishes for $\v_t^+\sim t^n=s^n(t/s)^n$ with $n=0,1,2$, we see that $\delta M_{s0}^{(-1,0)}\sim \dd\dd\dd \v^+_0$ and therefore the limit $\lim_{s\to0}\delta M_{s0}^{(-1,0)}$ is finite. Close to $s=1$ we have $M^{(0,0)}_{1s}\propto (1-s)^{-2\Delta}$ with $\Delta\in(0,1)$. Therefore, for $\Delta\ge1/2$ the integral is divergent, and we need to subtract the divergence using boundary terms in (\ref{eqn-Var-2ndOrd}). As before, we choose a keyhole subtraction scheme, where we deform the integral to go around the $s$-cut. Unlike the one variation integral (\ref{cutint}), the $s$-cut that starts at $s=1$ does not end at $s=0$. We denote the corresponding integral as
\beq\label{2-int}
\delta^2 M_{10}^{(0,0)} \Big|_{\text{2-int}}\equiv{\ii \eta \,|x_{10}|\over2\sin(2\pi\Delta)}\subset\hspace{-5mm}\int\limits_{(1,0}\dd s\,\v_s^- M_{1s}^{(0,0)}\delta M_{s0}^{(-1,0)}\,.
\eeq

Note that the order of integration is important -- in the scheme (\ref{2-int}) we first do the $t$-integral using the keyhole prescription and only then plug the result into the $s$-integral. Different choices of regularization of this double integral differ by the boundary term and can be absorbed into a shift in the coefficients $\gamma_i$ and $\tilde\gamma_i$.\footnote{We have verified this explicitly for several different schemes. See appendix \ref{apd:sec:2ndOrderSchemeInd} for details.} The regularization (\ref{2-int}) does not respect CPT and therefore in this scheme $\tilde\gamma_i\ne\gamma_i$.

\subsubsection{Summary}

In total, the second-order variation takes the form
\beq\la{delta2Mtotal}
 \delta^2 M_{10}^{(0,0)}=[\text{int}^2] +[\text{int}] + [\bar{B}B]+ [B^2] + [\bar{B}^2]\ ,
\eeq
where\footnote{Here, the tilded coefficients are chosen such that in a PT-symmetric scheme, they are equal to the un-tilded coefficients, $\gamma_i = \tilde{\gamma}_i$. The PT symmetry exchanges the left and right endpoints, exchanges the $\pm$ component, and introduces a negative sign to all derivatives. For instance, $\v_0^+ \dd \v_0^- \delta_3$ becomes $\v_1^- (-\dd) \v_1^+ (-\bar{\delta}_3) = \v_1^- \dd \v_1^+ \bar{\delta}_3$.} 
\begin{align}\la{summaryterms}
 [\bar{B}B]=&\v_1^+\v_0^-M_{10}^{(1,-1)} \,,\\
 [BB]=&\Big[ \gamma_0\v_0^+ \v_0^-\delta_3^2+\gamma_1\(\v_0^+ \dd \v_0^- + \v_0^- \dd \v_0^+\)\delta_3+\gamma_4\(\v_0^+ \dd \v_0^- - \v_0^- \dd \v_0^+\)\delta_3\nn\\
 &+\gamma_3 \dd \v_0^+ \dd \v_0^-+\gamma_2 \(\v_0^+ \dd \dd \v_0^- + \v_0^- \dd \dd \v_0^+\)+\gamma_5 \(\v_0^+ \dd \dd \v_0^- - \v_0^- \dd \dd \v_0^+\)\Big]M_{10}^{(0,0)}\,,\nn\\
 [\bar B\bar B]=&\Big[\tilde\gamma_0\v_1^+ \v_1^-\delta_3^2 +\tilde\gamma_1\(\v_1^+ \dd \v_1^- + \v_1^- \dd \v_1^+\)\delta_3 -\tilde\gamma_4\(\v_1^+ \dd \v_1^- - \v_1^- \dd \v_1^+\)\delta_3\nn\\
 &+\tilde\gamma_3 \dd \v_1^+ \dd \v_1^-+\tilde\gamma_2 \(\v_1^+ \dd \dd \v_1^- + \v_1^- \dd \dd \v_1^+\) - \tilde\gamma_5 \(\v_1^+ \dd \dd \v_1^- - \v_1^- \dd \dd \v_1^+\)\Big]M_{10}^{(0,0)}\,,\nn\\
[\text{int}]=&\,\Xi \,\int\dd s\(\dd \v_s^+\dd\dd \v_s^--\dd \v_s^-\dd\dd \v_s^+\)M_{10}^{(0,0)}\,,\nn\\
 [\text{int}^2]=&-\({\ii\eta |x_{10}|\over{2}\sin(2\pi\Delta)}\)^2\subset\hspace{-3.3ex}\int\limits_{(1,0)}\!\! \dd s\,\v_s^-  
M^{(0,0)}_{1s}\Big(\oint\limits_{[0,s]} \dd t\, \v_t^+M^{(-1,1)}_{st}M_{t0}^{(0,0)}\Big)\,.\nn
\end{align}
A change in the regularization scheme amounts to a change in the regularized double integral, $[\text{int}^2]$, together with a shift in the $\gamma_i$'s and $\tilde\gamma_i$'s coefficients, in such a way that the sum (\ref{delta2Mtotal}) is unaffected.

\subsubsection{The Normalization Independent Coefficients}

To compute the second-order deformation of the line, we need to fix the coefficients that appear in (\ref{summaryterms}). These are the $\gamma_i$'s, $\tilde\gamma_i$'s, $\eta$, $\Xi$, the overall normalization of $M^{(0,0)}$ and $M^{(-1,1)}$ that we have denoted by $c_0$ and $c_{-1}$ in (\ref{Mssb}), and $c_1$ in $[\bar{B}B]$. Among these, $\eta$, $c_0$, and $c_{-1}$ depend on the choice of normalization of the four bottom boundary operators in (\ref{bottom}). Therefore, they are not physical by themselves. However, the combination
\beq\la{Lambdadefinition}
\Lambda\equiv\eta^2c_0c_{-1}\,,
\eeq
is independent of the normalization of the boundary operators. This combination governs the two-point function of the displacement operator. It is given by\footnote{Instead, we can use the two-point function of $\mathbb D$ on $\llangle\lo^{(0,-1)}\ro^{(0,1)}\rrangle$, where it takes the form
\beq\la{DD2}
{\llangle\lo^{(0,-1)}{\mathbb D}_+(x_s){\mathbb D}_-(x_t)\ro^{(0,1)}\rrangle\over\llangle\lo^{(0,-1)}\ro^{(0,1)}\rrangle}=\Theta(s-t){\Lambda\over x_{st}^4}\({x_{10}x_{st}\over x_{1s}x_{t0}}\)^{2(2-\Delta)}\,.
\eeq
The same $\Lambda$ also governs the two-point function of the displacement operator on the circle, where it takes the more standard form 
\begin{equation}
{\<{\mathbb D}_+(x_0){\mathbb D}_-(x_\theta)\>_\text{circ}\over\<\One\>_\text{circ}}={\gamma\over[2R^2(1-\cos\theta)]^2}\,,\qquad\text{with}\qquad\gamma={\Lambda\over N\<\One\>_\text{circ}}\,,
\end{equation}
where $R$ is the radius of the circle and the $\pm$ components are with respect to the local transverse space. They are obtained from the straight-line ones by a conformal transformation.
}
\beq\la{DD}
{\llangle\lo^{(0,0)}{\mathbb D}_-(x_s){\mathbb D}_+(x_t)\ro^{(0,0)}\rrangle\over\llangle\lo^{(0,0)}\ro^{(0,0)}\rrangle}=\Theta(s-t){\Lambda\over x_{st}^4}\({x_{10}x_{st}\over x_{1s}x_{t0}}\)^{2\Delta}\,,
\eeq
where $\Theta$ is the step function. Here, we have divided the correlator by the expectation value of the same straight mesonic line operator without insertion of displacement operators. The normalization of the displacement operator is fixed by its physical interpretation in terms of the geometric displacement. Hence, this ratio is manifestly independent of our choice of normalization.

Another physical combination is the ratio
\beq
\Sigma\equiv{c_1/c_0
\over 2\Delta(2\Delta+1)}\,,
\eeq
where $c_1$ is defined in (\ref{c1definition}) and for convenience, we have factored out a factor of $2\Delta(2\Delta+1)$. Provided that $\ro^{(0,-1)}$ and $\ro^{(0,0)}$, and hence their normalizations, are related by the path derivative in (\ref{dm00}), this ratio is also physical. It can be thought of as the quantum version of the boundary equation of motion $\Box\phi=0$. To see this, consider the operator
\beq
\cM^{(0,-1)}\equiv\lo^{(0,0)}\cW\ro^{(0,-1)}\,,
\eeq
along a straight line. It is invariant under constant translation in the plus direction. Using (\ref{eqn-DeltaM1st}), (\ref{displacement}), (\ref{dpr}), (\ref{dpl}), and the fact that for a constant variation vector $\dd \v^\mu=\dd \dd \v^\mu=0$ in (\ref{boundartvar}), we end with
\beq\la{tovary}
\llangle\lo^{(0,1)}\cW\ro^{(0,-1)}\rrangle+\llangle\lo^{(0,0)}\cW\,\delta_+\ro^{(0,-1)}\rrangle=0\,.
\eeq
Following axiom \hyperlink{axiom3}{III}, the only operator with the same transverse spin and conformal dimension as that of $\delta_+\ro^{(0,-1)}$ is $\ro^{(2,0)}$. Hence, equation (\ref{tovary}) reduces to the relation
\beq \label{eqn-OneVariation-dOM1}
\delta_+\ro^{(0,-1)}=-\Sigma\,
\ro^{(2,0)}\,.
\eeq
In the free bosonic theory, this relation becomes $\d_+\d_-\phi=-{1\over2}\d_3^2\phi$. Similarly, applying constant translation in the minus direction to $\mathcal{M}^{(1,0)} \equiv \lo^{(1,0)} \mathcal{W} \ro^{(0,0)}$ implies 
\beq \label{eqn-OneVariation-dOM2}
\delta_- \lo^{(0,1)}=-\Sigma\,
\lo^{(2,0)}\,.
\eeq

In the following, we will use the axioms to bootstrap the $\gamma_i$' s, $\tilde\gamma_i$'s, $\Xi$, $\Lambda$, and $\Sigma$.

\subsection{Bootstrap}\la{bootstrapsec2nd}

\subsubsection{Constraints from Conformal Symmetry of a Straight Line}

Consider a general conformal transformation, $x_s\to\tilde x_{\beta}(x_s)$, that may shift the endpoints of the straight line only in the transverse direction. To linear order, the resulting deformation profile
\beq\la{straightconf}
x_\text{straight}\ \rightarrow\ x_\text{straight}+\v_c\,,
\eeq
is a quadratic polynomial in $s$
\beq \label{conftrans}
\v^+_c= \mathfrak{c}_0^+ + \mathfrak{c}_1^+ s + \mathfrak{c}_2^+ s^2\,, \qquad \v^-_c = \mathfrak{c}_0^- + \mathfrak{c}_1^- s + \mathfrak{c}_2^- s^2\,,
\eeq
with six independent coefficients; see Appendix \ref{sec:apd:confTransf} for more details. 

In the case of $M^{(0,0)}$, the conformal and spin factors in (\ref{source2}) take the form
\beq\la{transM00}
\begin{aligned}
&[\text{conformal factor}]=1-2\Delta\,\delta\log|x_{10}|+\dots\,,\\
&[\text{spin factor}]\qquad\ \,=1+{\mathfrak s}\(\delta\theta_0-\delta\theta_1\)
+\dots\,,
\end{aligned}
\eeq
with ${\mathfrak s}=(1/2-\Delta)$. For the conformal transformation (\ref{conftrans}) we find
\beq\la{eqn-2ndOrder-DistanceChange}
\begin{aligned}
\delta\log|x_{10}|&=\left.(\v_c^+)\right|_0^1\left.(\v_c^-)\right|_0^1/|x_{10}|^2=(\mathfrak{c}_1^-+\mathfrak{c}_2^-)(\mathfrak{c}_1^+ +\mathfrak{c}_2^+)/|x_{10}|^2\,,\\
(\delta\theta_0-\delta\theta_1)&={1\over2}\left.\(\v_c^+\dd\dd \v_c^--\v_c^-\dd\dd \v_c^+\)\right|_0^1=(\mathfrak{c}_1^-\mathfrak{c}_2^+-\mathfrak{c}_1^+\mathfrak{c}_2^-)/|x_{10}|^2
\,.
\end{aligned}
\eeq
It is therefore sufficient to keep the term linear in $\delta\log|x_{10}|$ and $\delta\theta$ in (\ref{transM00}) when working to second order in the deformation. Note that because $\dd\dd\dd \v_c=0$, the spin source is the same as the spinning integral $[\text{int}]$ in (\ref{summaryterms}), with $\Xi={\mathfrak s}/2$.

In our scheme, the double integral (\ref{2-int}) vanishes for a quadratic deformation profile (\ref{eqn-oneVar-KeyholeFinite}). Hence, the change in the expectation value (\ref{transM00}) comes only from the boundary terms. In particular, it cannot fix $\Lambda$ (\ref{Lambdadefinition}) because it is independent of $\eta$, (which only appears in $[{\rm int}^2]$ in (\ref{summaryterms})). By matching the coefficients in front of $\mathfrak{c}_0^\pm$, $\mathfrak{c}_1^\pm$ and $\mathfrak{c}_2^\pm$ between $[\bar{B}B]+ [B^2] + [\bar{B}^2]$ and (\ref{transM00}) we can reduce the set of independent unknown coefficients to $\{\gamma_0, \gamma_2, \gamma_4, \gamma_5,\Sigma,\Lambda\}$.

\subsubsection{Constraints from Conformal Symmetry of a Curved Line}
\label{sec:contraintsCurved2}

We see that the conformal symmetry of a straight line is not sufficient to fix all the coefficients that appear in the second-order variation (\ref{summaryterms}). Going to higher orders in the deformation does not improve the situation because the number of independent coefficients also grows. 
To determine the remaining coefficients, we demand that the expectation value of a mesonic line operator along an arbitrary smooth path, not necessarily straight, respects the conformal symmetry (\ref{source2}). 

Explicitly, we consider a path that we decompose into a straight line and an arbitrary smooth deformation
\beq\la{deformedpath}
x_a = x_{\mathrm{straight}} + \v_a\,.
\eeq
We then apply a conformal transformation to it,
\beq
x_a\ \rightarrow\ x_{ac}=x_a+\v_c=x_\text{straight}+\v_a+\v_c\,.
\eeq
While for a finite deformation, $\v_c$ depends on $\v_a$, at second order, they are independent. Therefore, we can simply expand $M^{(0,0)}[x(\cdot)+\v_a(\cdot)+\v_c(\cdot)]$ to second order using (\ref{delta2Mtotal}). The term of order $O(\v_c^2)$ was considered in the previous section. The term of order $O(\v_a^2)$ is not constrained because $\v_a$ is arbitrary. We then focus on the term of order $O(\v_a\v_c)$ and demand that it is consistent with the conformal transformation of the deformed path (\ref{deformedpath}). As before, it is sufficient to have both $\v_a$ and $\v_c$ transverse at the boundaries of the straight line. 

Because the line is not straight before the conformal transformation, the source term, due to the boundary conformal weight (\ref{transM00}), can have a support that is linear in $\v_c$. For the same reason, the boundary polarization vector $n^\text{int}$ in (\ref{source2}) is not a vector in the transverse plane of the straight line. Hence, to expand the mesonic lines around the straight line, we now have to further rotate the boundary polarization vectors. After doing so, we can compare the expansions of the numerator and denominator in (\ref{source}) around the straight line. Starting with $n$ and $\tilde n$ in (\ref{source}) on either of the two sides of the line, this results in the following sequence of rotations
\beq
\tilde n\ \rightarrow\ \tilde n^\text{straight}\ \rightarrow\ n^\text{straight}\ \rightarrow\ n\,,
\eeq
where $n\rightarrow n'$ stands for a rotation of the polarization vector $n$ to the polarization vector $n'$ in the plane that these two vectors span. This sequence of rotations results in the spin factor source
\beq\la{spinfactorac}
[\text{spin factor}]_{ac}=\left.{\mathfrak s}\[\(\v_a^+\dd\dd \v_c^--\v_a^-\dd\dd \v_c^+\)+\(\dd \v_c^+\dd \v_a^--\dd \v_c^-\dd \v_a^+\)/2\]\right|_\text{boundary}\,.
\eeq

For definiteness, we take the deformed path to be a degree-four polynomial
\beq
\v_a(s)=\sum_{n=0}^4{\mathfrak a}_n^\pm s^n\,.
\eeq
Consequently, we require that the second-order variation (\ref{delta2Mtotal}) agrees with the spin and conformal factors for any value of $\mathfrak{c}_{n=0,1,2}^\pm$, and $\mathfrak{a}^\pm_{n=0,\ldots,4}$. This requirement, together with the constraint obtained at order $\v_c^2$, fixes all coefficients and the normalization independent quantities to
\beq \label{eqn-2ndOrder-AllValueGamma}
\begin{aligned}
\gamma_1=&\gamma_3=-\gamma_4=-\tilde\gamma_3=(1/2-\Delta)/2\,,\\
\gamma_2=&-\gamma_5=(\Delta-1)\gamma_1\,,\\
\tilde\gamma_1=&\tilde\gamma_4=2\tilde\gamma_2=2\tilde\gamma_5=-\gamma_0=1/2\,,\\
\tilde\gamma_0=&0\,,
\end{aligned}
\eeq
and 
\beq \label{LambdaXiSigma}
\Lambda= -\frac{1}{2 \pi }(2\Delta -1) (2 \Delta -2) (2 \Delta -3) \sin(2 \pi \Delta)\,,\quad\Xi=0\,,\quad\Sigma=1/2\,.
\eeq

Note that $c_1\propto\Sigma\ne0$, (\ref{LambdaXiSigma}). Hence, the operators $\ro^{(0,-1)}$ with dimension given in (\ref{Deltam1}) must exist. We see that it is required for the consistency of $M^{(0,0)}$ with conformal symmetry. If instead of the second-order variation of $M^{(0,0)}$ we consider that of $M^{(-1,1)}$, we conclude that the primary boundary operator $\ro^{(0,2)}$ must also exist, with $\Delta^{(0,2)}=\Delta^{(0,1)}+1=3-\Delta$. The two derivations are related by the tower-swap discussed below (\ref{spectrumR}).

In general, we can use polynomials of arbitrarily high degree as the basis for smooth path deformations. Every power of $s$ in this deformation leads to a new constraint. For the consistency of the solution (\ref{eqn-2ndOrder-AllValueGamma}), (\ref{LambdaXiSigma}), all these constraints must be satisfied. By considering polynomials of high degree, we have verified that this is indeed the case. At the third and higher order in the deformation of the straight line, there are more and more coefficients to fix. Correspondingly, in these orders, it is necessary to consider polynomial deformations of a higher degree and, in that way, generate more and more independent constraints.

\subsubsection{Two Towers of Spinning Boundary Operators}

One of the results of the previous section is the existence of the primary boundary operators, $\ro^{(0,-1)}$ and $\ro^{(0,2)}$. Using these operators, we can now derive the existence of two towers of primary boundary operators with increasing transverse spin, (\ref{spectrumR}). 

Consider, for example, the second-order variation of the operator $M^{(1,-1)}$. The main difference from the second-order variation of $M^{(0,0)}$ is that these boundary operators do not overlap with the displacement operator. Hence, the second-order variation of the operator $M^{(1,-1)}$ does not contain a double integral. It does, however, contain a single integral. The structure of this single integral is the same as in Section \ref{sec:2ndOrder:singleInt}. That is, it is proportional to the insertion of the identity on the line or comes with a (positive) fractional power of $\epsilon$. Therefore, the only finite terms in this case are
\beq\la{delta2M1m1}
 \delta^2 M_{10}^{(1,-1)}=[\text{int}] + [\bar{B}B]+ [B^2] + [\bar{B}^2]\,.
\eeq
Using integration by parts, the single integral can be written as
\beq\la{spinningint}
[\text{int}]=\Xi^{(1,-1)} \,\int\dd s\(\dd \v_s^+\dd\dd \v_s^--\dd \v_s^-\dd\dd \v_s^+\) M^{(1,-1)}_{10}\,.
\eeq
Here, we have introduced the new parameter $\Xi^{(1,-1)}$ because, in principle, it could be different from $\Xi=0$. The boundary variation terms represent
\beq
[\bar{B}B]=\llangle \delta\lo^{(0,1)} \delta  \ro^{(0,-1)} \rrangle\,,\quad [B^2]=\llangle \delta^2 \lo^{(0,1)} \ro^{(0,-1)} \rrangle \,,\quad [\bar{B}^2]=\llangle \lo^{(0,1)} \delta^2  \ro^{(0,-1)} \rrangle\,.
\eeq
They consist of all finite terms that are allowed by symmetry. First, we have
\beq\la{expd2O1}
\begin{aligned}
\delta^2 \lo^{(0,1)} & = \v_1^+  \v_1^- \bar{\beta}_{0,0} \lo^{(2,1)} + \dd \v_1^+  \v_1^- \bar{\beta}_{1,0} \lo^{(1,1)} + \v_1^+ \dd  \v_1^- \bar{\beta}_{0,1} \lo^{(1,1)} + \dd \dd \v_1^+  \v_1^- \bar{\beta}_{2,0} \lo^{(0,1)} \\
& + \v_1^+ \dd \dd \v_1^- \bar{\beta}_{0,2} \lo^{(0,1)} + \dd  \v_1^+ \dd \v_1^- \bar{\beta}_{1,1} \lo^{(0,1)} + \cdots\,. 
\end{aligned}
\eeq
Here, we have introduced six unknown coefficients $\bar{\beta}$, whose indices label the number of path derivatives that act on each of the variation vectors, $\bar{\beta}_{a,b} \dd^a \v_1^+ \dd^b \v_1^-$, with $\dd^0 v=v$, $\dd^1 v=\dd\v$, etc. The expansion of $\delta^2 \ro^{(0,-1)}$ takes the same form, with the replacements $\v_1 \mapsto \v_0$, $\bar{\beta} \mapsto \beta$, and $\lo \mapsto \ro$.

Next, we have the boundary operator
\beq\label{eq:deltaO}
\delta\ro^{(0,-1)} = \v_0^-\delta_-\ro^{(0,-1)}- \Sigma \v_0^+ \ro^{(2,0)} + \beta_1 \dd \v_0^+ \ro^{(1,0)} + \beta_2 \dd \dd \v_0^+ \ro^{(0,0)} + ...
\eeq
where $\Sigma={1\over2}$ was defined in (\ref{eqn-OneVariation-dOM1}) and determined in (\ref{LambdaXiSigma}). For the operator $\delta_-\ro^{(0,-1)}$ to give a finite contribution, it must have dimension $\Delta^{(0,-2)}=\Delta^{(0,-1)}+1=\Delta+2$. We denote it by $\ro^{(0,-2)}=\delta_-\ro^{(0,-1)}$, if it exists. 

Similarly, the expansion of $\delta_- \lo^{(0,1)}$ is obtained from (\ref{eq:deltaO}) by replacing $\v_1 \mapsto \v_0$, $\bar{\beta} \mapsto \beta$, and $\lo \mapsto \ro$. Putting them together we get,
\begin{align}
\llangle \delta\lo^{(0,1)} \delta\ro^{(0,-1)} \rrangle =&\ \v_1^- \v_0^+M_{01}^{(2,-2)}\\
&+\Big({1\over2}\big[{\beta}_1 \v_1^- \dd \v_0^+ \bar{\delta}_3^2 \delta_3  - \bar{\beta}_1\dd \v_1^- \v_0^+ \bar{\delta}_3 \delta_3^2+ {\beta}_2 \v_1^- \dd \dd \v_0^+ \bar{\delta}_3^2 + \bar{\beta}_2 \dd \dd \v_1^- \v_0^+ \delta_3^2\big]\nn\\
&  +\bar{\beta}_1 {\beta}_1 \dd \v_1^- \dd \v_0^+ \bar{\delta}_3 \delta_3  + \bar{\beta}_2 {\beta}_1 \dd \dd \v_1^- \dd \v_0^+ \delta_3 + \bar{\beta}_1 \beta_2\dd \v_1^- \dd \dd \v_0^+ \bar{\delta}_3\nn\\
& + \bar{\beta}_2 \beta_2 \dd \dd \v_1^- \dd \dd \v_0^++{1\over4}\v_1^- \v_0^+ \delta_3^2 \Big) M_{01}^{(0,0)}\,,\nn
\end{align}
where 
\beq
M_{01}^{(2,-2)}= \llangle \lo^{(0,2)} \ro^{(0,-2)} \rrangle =\frac{c_2}{|x_{10}|^{2(\Delta+2)}}\, .
\eeq

The primary boundary operators $\ro^{(0,-1)}$ and $\lo^{(0,1)}$ have dimension $\Delta+1$ and spins ${\mathfrak s}=\pm(1/2+\Delta)$. Hence, a mesonic line that ends on these operators transforms under a conformal transformation as
\beq\la{deltaM1m1}
{\delta^2 \<\lo^{(0,1)}\cW\ro^{(0,-1)}\>\over\<\lo^{(0,1)}\cW\ro^{(0,-1)}\>}=
- 2 (1+\Delta) \, \delta \log |x_{10}|+\delta[\text{spin source}]\,,
\eeq
where the spin source at order $\v_c^2$ is given in (\ref{transM00}) and at order $\v_a\v_c$ in (\ref{spinfactorac}). 
The deformation (\ref{deltaM1m1}) must reproduce these sources. By choosing a generic polynomial $\v_a$ and equating the terms of order $\cO(\v_c^2)$ and $\cO(\v_a \v_c)$ on the two sides, we find a unique solution. \footnote{In practice, we find that choosing $\v_a^\pm =  \mathfrak{a}_0^\pm s^3 + \mathfrak{a}_1^\pm s^4 + \mathfrak{a}_2^\pm s^5$ to be a fifth-order polynomial in $s$ determines the answer completely.} The resulting coefficients are compatible with the CPT symmetry
\beq
\beta_{i,j} = \bar{\beta}_{j,i}\,, \qquad \beta_{i} = \bar{\beta}_{i}\,, \qquad \forall i,j \, .
\eeq
They are given by 
\beq\la{betas}
\begin{aligned}
c_2 & =  (\Delta +1) (2 \Delta +3) c_1 =\Delta  (\Delta +1) (2 \Delta +1) (2 \Delta +3)c_0\,, \\
\beta_{1,0}&=\beta_1=2\beta_{1,1}= -(2\Delta+1)/2\,,\qquad \Xi^{(1,-1)}=0 \,,\\
\beta_{0,0} & = -1/2\,, \qquad \beta_{0,1} = \beta_{0,2} = 0\,,\qquad\beta_{2,0} =\beta_2=\Delta\beta_1\,.
\end{aligned}
\eeq

Because in this case, there is no integral to regularize, all of these coefficients are scheme-independent. In particular, the values of $\beta_{0,0} = \bar{\beta}_{0,0} = -1/2$ can be interpreted as the quantum version of the equation of motion with one extra $\d_+$ derivative, $\partial_+ \partial_-(\d_+\phi)= -\frac{1}{2} \partial_3^2(\d_+\phi)$.

Since $c_2\ne0$, the operators $\lo^{(0,2)}$ and $\ro^{(0,-2)}$, of conformal dimension $\Delta^{(0,-2)}=\Delta+2$ must exist. We can repeat the same computation successively, starting with $M^{(-\ell,\ell)}$ for all $\ell \leq-2$. They are related to the case of $M^{(1,-1)}$ by substituting $\Delta \rightarrow \Delta+|\ell|$, and $\{c_0, c_1, c_2\}  \rightarrow \{c_{-\ell-1}, c_{-\ell}, c_{1-\ell}\}$ into the above bootstrap. In this way, we deduce the existence of infinite towers of operators of the form
$\lo^{(0,-\ell)}$ and $\ro^{(0,\ell)}$ with $\ell\le-1$, whose spectrum is given by (\ref{spectrumR}). They are subject to the operator equation
\beq
\delta_- \delta_+ \lo^{(0,\ell)} = -\frac{1}{2} \lo^{(2,\ell)}\, , \quad \delta_+ \delta_- \ro^{(0,\ell)} = -\frac{1}{2} \ro^{(2,\ell)}\,.
\eeq
Correspondingly, the expectation value of $M^{(\ell,\ell')}$ is given by (\ref{Mssb}), with 
\beq
c_{\ell + 1} = (2\Delta+2\ell +1) (\Delta+\ell ) c_\ell=c_0 \prod_{j=0}^\ell (2\Delta+2j +1) (\Delta+j)\,,\qquad \ell\ge0 \, .
\eeq

Similarly, we deduce the existence of infinite towers of operators of the form
$\lo^{(0,-\ell)}$ and $\ro^{(0,\ell)}$ with $\ell\ge1$, whose spectrum is given by (\ref{spectrumR}). Their normalizations are related to $c_{-1}$ as
\beq
c_{-\ell-1}=(1+\ell-\Delta)(3+2\ell-2\Delta)\,c_{-\ell}=c_{-1}\prod_{j=1}^{\ell}(1+j-\Delta)(3+2j-2\Delta)\,,\qquad \ell\ge1\,.
\eeq
Their derivation is related to that above by tower-swap, see discussion below (\ref{spectrumR}).

\subsubsection{Bootstrap of Spinning Mesonic Lines}
We can also apply the bootstrap to operators that have a non-zero transverse spin and, therefore, a zero expectation value. As an example, let us examine the operators $M^{(1,1)}_{10}$ and $M^{(0,2)}$. These operators have two units of transverse spin, ${\mathfrak s}+\bar{\mathfrak s}=2$. Hence, to absorb that spin and have a non-zero expectation value, we need at least two variations in the plus direction.

\paragraph{The Operator $M^{(1,1)}_{10}$.}
The only finite contributions to the second-order variation of this operator have one bulk integration. They take the form 
\beq \label{eqn-2ndOrder-VarOpM1M1}
\delta^2 M^{(1,1)}_{10} =\int\limits_0^1 \! \dd s\ \Big[\v^-_1 \v^-_s \llangle {\delta}_-\lo^{(0,1)}\mathbb{D}_{-}(x_s) \ro^{(0,1)}\rrangle+\(\v^-_s\)^2\llangle \lo^{(1,0)}\mathbb{D}_{--}(x_s)  \ro^{(0,1)}\rrangle \Big]_\text{reg}\,,
\eeq
where $\left[\dots\right]_\text{reg}$ stands for a regularized integral. The boundary variation ${\delta}_- \lo^{(0,1)}$, appearing in the second term, was defined below \eqref{eq:deltaO}. The coefficients in \eqref{eq:deltaO} are scheme independent and were fixed in (\ref{betas}). The line operator $\mathbb{D}_{--}$ is some linear combination of primary operators of dimension $3$ and spin $2$. Among these, only $\ro^{(0,-1)} \lo^{(0,-1)}$ can contribute to (\ref{eqn-2ndOrder-VarOpM1M1}). We denote its coefficient by $\eta^{(-1,-1)}$,
\beq
\mathbb{D}_{--}=\eta^{(-1,-1)}\[\ro^{(0,-1)}\times\lo^{(0,-1)}\](2n_-^2)+\dots\,,
\eeq
where the dots stand for operators that do not overlap with either $\ro^{(0,1)}$ on the right end of a straight line or with $\lo^{(0,1)}$ on the left end. 

We proceed as before, picking a sufficiently generic deformation $\v_a$ and applying a conformal transformation to it. Because the operator $\cM^{(1,1)}$, and its first-order variation $\delta\cM^{(1,1)}$ have zero expectation value, so do their conformal transformations. Consequently, at second order, the terms $\cO(\v_a \v_c)$ and $\cO(\v_c^2)$ on the right-hand side of \eqref{eqn-2ndOrder-VarOpM1M1} must independently sum to zero.

We have used the keyhole regularization scheme to evaluate the two integrals in (\ref{eqn-2ndOrder-VarOpM1M1}), which are, however, scheme independent. The vanishing of the $\cO(\v_a \v_c)$ contribution fixes
\beq\la{eta11}
\eta^{(-1,-1)} = \frac{1}{2}\,.
\eeq

It is worth mentioning that although the values of $\beta_1$ and $\beta_2$ were found in the preceding section, we can also fix them using the bootstrap equation used in this section. Because these are coefficients of local operators on the line, the result is, of course, consistent with what we have found before. The order in which one solves the bootstrap constraints is a matter of preference. 

\paragraph{The Operator $M^{(0,2)}_{10}$.}

Similarly, for this operator we have
\beq \label{eqn-2ndOrder-VarOpM2Op0}
\begin{aligned}
\delta^2 M^{(0,2)}_{10} & =\int\limits_0^1 \!\! \dd s\ \Big[\v_s^-\v_0^-\llangle \lo^{(0,0)}\mathbb{D}_-(x_s) \delta_-\ro^{(0,2)}\rrangle + (\v_s^-)^2\llangle \lo^{(0,0)}\mathbb{D}_{--}(x_s)\ro^{(0,2)}\rrangle \Big]_\text{reg}\, ,  \\
\end{aligned}
\eeq
Because of the different boundary spin assignment, we can now probe the coefficient of $\ro^{(0,0)} \lo^{(0,-2)}$ in $\mathbb{D}_{--}$,
\beq
\mathbb{D}_{--}=\eta^{(0,-2)}\[\ro^{(0,0)}\times \lo^{(0,-1)}\](2n_-^2)+\dots\, .
\eeq

The expansion of $\delta_-\ro^{(0,2)}$ in terms of boundary operators takes the form 
\beq
    \delta_- \ro^{(0,2)} = - \v_0^- \frac{1}{2} \ro^{(2,1)} + {\beta}_1 \dd \v_0^- \ro^{(1,1)} + {\beta}_2 \dd \dd \v_0^- \ro^{(0,1)} + ...\,.
\eeq
We find
\beq
    {\beta}_1 =\Delta-\frac{5}{2}\,, \quad {\beta}_2 = (2-\Delta){\beta}_1\,, \qquad \eta^{(0,-2)} = \frac{1}{2} \,. 
\eeq

This result can also be obtained simply by interchanging $\Delta \to 2-\Delta$ in the deformation of $M^{(-1,-1)}$. That is due to the invariance of the spectrum under tower-swap.

To summarize, at second order, we always find a unique bootstrap solution. It can now be used to explicitly compute $\delta^2M$ for an arbitrary smooth deformation profile.

\section{Third Order}\la{sec-third order}

We can apply the bootstrap method developed in the previous sections to higher-order deformations. At each order, we pick a regularization scheme, list all the finite terms that can appear, and fix their coefficients by imposing the conformal symmetry of a curved line. All the ingredients that appear in higher orders are already present at the third order. In this section, we carry out the method in third order explicitly and discuss the higher-order generalizations. Our main point in doing so is to demonstrate that the bootstrap solution is unique and that it allows one to systematically compute the expectation value of mesonic lines to any order.

\subsection{Line Variation}

The third-order variation of any mesonic line can be divided into boundary and bulk variations as follows
\beq \label{eqn-Var-3rdOrd}
\begin{aligned}
\delta^3\cM & =\delta^3 \bar{O}\,\mathcal{W}\, O  +\delta^2\,\bar{O}\,\mathcal{W}\,\delta O +\bar{O}\, \mathcal{W}\,\delta^3 O +\delta \bar{O}\,\mathcal{W}\,\delta^2 O\, \\
&  +\delta\bar{O}\,\delta\mathcal{W}\,\delta O +\delta^2\,\bar{O}\,\delta\mathcal{W}\, O+\bar{O}\, \delta\mathcal{W}\,\delta^2 O\, \\
& +\bar{O}\,\delta^2\mathcal{W}\,\delta O+ \delta \bar{O}\,\delta^2\mathcal{W}\, O+\bar{O}\,\delta^3\mathcal{W}\,O  \,.
\end{aligned}
\eeq
As before, these expressions encode many contributions. In particular, the third-order variation at the bulk of the line, $\delta^3\cW$, includes the integration of three displacement operators
\beq\la{eqn-Var-bulk-3Var}
[\text{int}^3]_\text{bare}=|x_{10}|^3 \int\limits_0^1\dd s\,\v_s^\mu\int\limits_0^s\dd t\,\v_t^\nu\int\limits_0^t\dd u\,\v_u^\rho\,\llangle\lo\,{\mathbb D}_\mu(x_s){\mathbb D}_\nu(x_t){\mathbb D}_\rho(x_u)\ro\rrangle\,.
\eeq

The transverse spin of a third-order deformation is either one or three. Hence, to have a non-zero variation, we need to start with a spinning straight mesonic line. 
In the following, we have chosen to present the third-order variation of $M^{(0,1)}_{10}$.

Local, scheme-independent terms in $\delta^3\cM$ involving one or two variations are the same as they are at second order. For example, we have 
\beq
\begin{aligned}
\<\delta\bar{O}^{(0,0)}\,\delta^2\mathcal{W}\, O^{(0,1)}\>&=\v_1^{+}\int\dd s(\v_s^-)^2\llangle \lo^{(0,1)}\mathbb{D}_{--}(x_s) \ro^{(0,1)}\rrangle+\dots\\
&=\v_1^{+}\eta^{(-1,-1)}\int\dd s\, (\v_s^-)^2 \, M_{1s}^{(1,-1)} M_{s0}^{(-1,1)}+\dots\,,
\end{aligned}
\eeq
where $\eta^{(-1,-1)}$ 
was determined in (\ref{eta11}). If we use the same regularization scheme that we have used at second order, then other local scheme-dependent terms would also be the same. For that purpose, we first regularize the triple integral in (\ref{eqn-Var-bulk-3Var}) by converting the middle integral into a contour integral using \eqref{keyhole3}. We then use the keyhole prescription to regulate the remaining right and left integrals\footnote{To regularize the double integral contribution to $\delta^2M_{10}^{(0,0)}$ we have used the scheme of \eqref{2-int}. Similarly, for the double integral in $\delta^2M_{10}^{(-1,1)}$ we use
\beq\label{2-int--+}
\delta^2 M_{10}^{(-1,1)} \Big|_{\text{2-int}}\equiv{\ii \eta \,|x_{10}|\over2\sin(2\pi\Delta)}\supset\hspace{-4.5mm}\int\limits_{1,0)}\dd s\,\v_s^- \delta M_{1s}^{(-1,0)} M_{s0}^{(-1,1)}\,.
\eeq
}
\beq 
[\text{int}^3]=
\({\eta|x_{10}|\over2 \ii\sin(2\pi\Delta)}\)^2 \subset\hspace{-5mm}\int\limits_{(1,0}\dd s\,\v_s^-\supset\hspace{-5mm}\int\limits_{s,0)}\dd u\,\v_u^-\,M^{(0,0)}_{1s}\,\delta M^{(-1,0)}_{su}\,M^{(-1,1)}_{u0}\,,
\eeq
where the order of the last $s$ and $u$ integrals does not matter.\footnote{This occurs because, after the middle integral, the integrals over $s$ and $u$ factor out into two separate integrals: one depends only on $s$, and the other only on $u$.}

In this regularization scheme, which is compatible with our second-order variation, the only new coefficients are those that multiply finite terms with three variations at the same point. All such terms have a single integral. They take the form
\begin{align}\la{intmid}
\text{[int]}_\text{mid}=&\frac{\eta |x_{10}| }{2 \ii \sin(2 \pi \Delta)}\oint\dd s\,M_{1s}^{(0,0)}\left[\chi_0\,\v^+_s(\v^-_s)^2\delta_3^2+\(\chi_1\,(\v_s^-)^2\dd\v_s^++\chi_2\,\v_s^+\v_s^-\dd\v_s^-\)\delta_3\right.\\
&+\left.\(\chi_3\,\v_s^-\dd\v_s^+\dd\v_s^-+\chi_4\,\v_s^+(\dd\v_s^-)^2+\chi_5\,(\v_s^-)^2\dd\dd\v_s^++\chi_6\,\v_s^-\v_s^+\dd\dd\v_s^-\)\right]M_{s0}^{(-1,1)}\,,\nn
\end{align}
where, for convenience, we have used integration by parts to put all the longitudinal operator derivatives on the right.\footnote{The boundary terms that result from these i.b.p. are counter-terms and therefore do not produce finite contributions.} We have regularized the divergences by converting the integral into a contour integral and absorbing all factors into the $\chi$'s. 

Altogether, the third-order variation takes the form
\begin{align}\la{eq:var3rd}
\delta^3 M^{(0,1)}_{10}=&\frac{\eta |x_{10}| }{2 \ii \sin(2 \pi \Delta)}\oint\limits_{[0,1]} \dd s\, \v_s^- \, \Big[M_{1s}^{(0,0)} \([BB]^{(-1,1)}_{s0}+\eta^{(0,-2)}\v_s^-\v_0^+M_{s0}^{(-2,2)}\)\\
&\qquad\qquad\qquad\qquad\qquad\quad\ +\([\bar B\bar B]_{1s}^{(0,0)}+\eta^{(-1,-1)}\v_s^-\v_1^+\,M_{1s}^{(1,-1)}\) M_{s0}^{(-1,1)}\Big]\nn\\
&+ [\text{int}]_\text{mid}+[\text{int}^3]\,,\nn
\end{align}
where $[\bar B\bar B]^{(0,0)}_{1s}$ is given in (\ref{summaryterms}) and is evaluated along $x_{1s}$ instead of $x_{10}$. $[BB]_{s0}^{(-1,1)}$ can be obtained from $[\bar B\bar B]^{(0,0)}_{1s}$ using tower-swap and CPT, see the discussion below (\ref{spectrumR}) and (\ref{2-int--+}). 
In practice, $[BB]_{s0}^{(-1,1)}$ is related to $[BB]_{s0}^{(0,0)}$ by first exchanging the $\gamma$'s with the corresponding $\tilde{\gamma}$'s, then flipping the signs of $\gamma_{4,5}$ and $\tilde{\gamma}_{4,5}$, and finally replacing $\Delta\to2-\Delta$.

\subsection{Bootstrap}

To fix the coefficients $\chi_i$ in (\ref{intmid}) we impose the consistency of (\ref{eq:var3rd}) with conformal symmetry. That is, we impose that the mesonic line transforms covariantly under conformal transformations $x\to\tilde x_\beta(x)$. As before, we choose the conformal transformations such that they move the endpoints in the transverse directions. Recall that $\beta$ parameterizes a family of conformal transformations that continuously interpolate between $x=\tilde x_0(x)$ and $\tilde x_\beta(x)$. Unlike the previous orders, at third order, the expansion in the deformation of the path and the expansion in powers of $\beta$ are not the same. 

Consider first the conformal transformation of a straight line, (\ref{straightconf}). The conformal deformation profile $\v_c$ can be expanded as
\beq
\v_c=\v^{(1)}_c + \v^{(3)}_c + O(\beta^4)\,,
\eeq
where $\v^{(n)}_c\propto\beta^n$, see Appendix \ref{sec:apd:confTransf} for explicit expressions. Because $M^{(0,1)}$ 
has zero expectation value, the constraint coming from the conformal variation of a straight line reads,
\begin{equation}
    \delta^3_{\v^{(1)}_c} M^{(0,1)}_{10} + \delta^1_{\v^{(3)}_c} M^{(0,1)}_{10} = 0\,,
\end{equation}
where $\delta^1_{\v^{(3)}_c} M^{(0,1)}_{10}$ is the first order variation (\ref{keyhole3}), evaluated with the profile $\v^{(3)}_c$. Similarly, $\delta^3_{\v^{(1)}_c} M^{(0,1)}_{10}$ is the third-order variation (\ref{eq:var3rd}), evaluated with the profile $\v^{(1)}_c$.

If instead of the straight line, we apply a conformal transformation to a deformed line (\ref{deformedpath}), then the conformal deformation profile admits a double expansion, in $\beta$ and in $\v_a$,
\beq
\v_c=\v^{(1,0)}_c + \v^{(3,0)}_c+\v_c^{(1,2)} +  \v_c^{(2,1)} + \dots\,,
\eeq
where $\v_c^{(n,m)}\propto\beta^n\v_a^m$, see Appendix \ref{sec:apd:confTransf} for explicit expressions. 

We now impose the conformal covariance of $M^{(0,1)}$ at orders $\beta\v_a^2$ and $\beta^2\v_a$. The conformal and spin sources are proportional to the mesonic line expectation value before the conformal transformation. The latter starts at order $\v_a$. Hence, the corresponding bootstrap constraint reads
\beq \label{thirdordereq}
\begin{split}
    &\(\delta^3_{\v^{(1)}_c+\v_a}-\delta^3_{\v_a} + \delta^1_{\v^{(3)}_c+\v_c^{(2,1)}+\v_c^{(1,2)}}\) M^{(0,1)}_{10} \\
    =&\([\text{conformal factor}]+[\text{spin factor}]\) \delta^1_{\v_a} M^{(0,1)}_{10}\,.
\end{split}
\end{equation}

The terms in the bracket on the RHS all starts at order $
\cO(\beta^2)$ and $\cO(\beta\v_a)$. The conformal factor reads
\beq
[\text{conformal factor}]=-2\Delta\, \delta\log|x_{10}|\,,
\eeq
where the change in distance is evaluated using $\v_c^{(1)} +\v_a$.
The spin factor
is given by
\begin{equation}
    [\text{spin factor}]= \frac{1}{2} \(\mathfrak{s}S_0+\mathfrak{s}'S_1\)\,,
\end{equation}
where $S=S^{(1,1)}+S^{(2,0)}$ and 
\beq 
\begin{aligned}
S^{(1,1)}=&\,2(\dd \dd\v_c^{(1)})^- \v^+_a - 2(\dd \dd \v_c^{(1)})^+ \v^-_a+(\dd\v_c^{(1)})^-\dd \v^+_a - (\dd \v_c^{(1)})^+\dd \v^-_a\,,\\
S^{(2,0)}=&\,(\v_c^{(1)})^+ (\dd \dd \v_c^{(1)})^--(\v_c^{(1)})^- (\dd \dd \v_c^{(1)})^+\,.
\end{aligned}
\eeq

By implementing (\ref{thirdordereq}) in \verb"Mathematica" and using a polynomial of degree four or higher for $\v_a$ we obtain a unique and consistent solution for the $\chi$'s. They are given by
\beq\la{chis}
\chi_1= \chi_2 = 0\,,\qquad\chi_0= \frac{\chi_3}{2} = \chi_4 =2 \chi_5 =\chi_6= -\frac{1}{2}\, .
\eeq

\subsection{Higher Orders}

The third-order analysis is sufficient to demonstrate the general structure. We work recursively in the order of the deformation. At the $n$'th order, we list all finite terms that can contribute to the mesonic line expectation value. Due to the locality of the line effective action, the coefficients of most of these terms are determined from lower orders, provided that we keep using the same regularization scheme. The only new terms have all the $n$ variations at the same point. This point can either be integrated along the line or at the boundary.\footnote{In the case of the third-order deformation of $M^{(0,1)}$, the terms with the three variations at the same boundary point are either irrelevant or that they are divergent counter-term.} To fix these coefficients, we impose that the deformed line, along $x_a$ (\ref{deformedpath}), transforms covariantly under conformal transformations. For that aim, we expand $\tilde x_\beta(x_a)$ in powers of $\beta$ and $\v_a$, keeping the sum of the two powers equal to $n$. We then collect all terms with fixed power $\beta^m\v_a^{n-m}$ in the variation of the straight line. These receive contributions from all orders in the line variation, up to order $n$. We equate the result to the expansion of the spin source and the conformal source at order $\beta^m\v_a^{n-m}$.

By considering a polynomial deformation profile, $\v_a$, we generate a sufficient set of equations to fix the solution. These equations are all linear in the new coefficients. Generating more equations serves as a consistency check for the solution. Once all coefficients are fixed, we can go back and evaluate the $n$'th order line variation explicitly.

\section{Discussion}\la{sec-discussion}

In this paper, we have developed a bootstrap method for studying conformal line defects. We have found that imposing the conformal symmetry of line defects along an arbitrary smooth path leads to new nontrivial functional constraints. Focusing on line defects in CS-matter theories at large $N$, we have observed that these constraints lead to a unique bootstrap solution and are therefore sufficient to fix the defect expectation value. The following is a list of future directions that we find interesting. 

The one-dimensional defect CFT that we considered is relatively simple. This is largely due to the factorization of the correlators on the line. In particular, we used it to compute the correlation functions between $n$ insertions of the displacement operator that arise at the $n$'th order of the line deformation. In other theories, with more complicated line defects, such correlators are unknown functions of the conformal cross-ratios. It would be interesting to see if these functions can be bootstrapped in a similar way and used to compute the expectation value of the defect. What makes us optimistic is that the smooth line deformation profile, $\v_a(s)$, entering the bootstrap constraints, is arbitrary. Therefore, it generates a functional bootstrap equation that can be used to gradually localize the individual deformation points and fix the $n$-point correlation of the displacement operator.

Large $N$ CS-matter theories are expected to be holographically dual to parity-breaking versions of Vasiliev's higher-spin theory, \cite{Vasiliev:1992av,Giombi:2011kc,Aharony:2011jz,Aharony:2012nh,Klebanov:2002ja,Sezgin:2003pt,Chang:2012kt,Leigh:2003gk}. The massless high-spin spectrum of this theory suggests that a unified description in terms of a tensionless, partly topological string may exist. In particular, it is not known how to incorporate the mesonic lines into Vasiliev's higher-spin theory. We expect that doing so will require a stringy description and that the mesonic lines may even be instrumental in constructing it.

The bootstrap method that we have employed gives the defect expectation value in an expansion around the straight line. It would be interesting to resum this expansion and find a more explicit representation of the conformal invariant functional of the path $F[x(\cdot)]$ in (\ref{generalform}). We expect the holographic description of the mesonic line to be such a representation. 

Finally, CS-matter theories possess a higher spin symmetry that is broken only at order $1/N$. An implication of this symmetry is the existence of protected primary tilt operators on the line, \cite{Gabai:2022mya}. It would be interesting to fix these operators and use them to derive bootstrap constraints in the same way that we have used the conformal symmetry. These high-spin constraints are expected to become important at the next order in $1/N$, while at the planar order at which we are working, they are redundant.

\acknowledgments
We thank Nathan Agmon, Ofer Aharony, Nissan Itzhaki, Zohar Komargodski, Yaron Oz, Yifan Wang, and Shimon Yankielowicz for useful discussions. AS thanks the Simons Center for Geometry and Physics for its hospitality. AS and DlZ are supported by the Israel Science Foundation (grant number 1197/20). During different periods of this work, BG was partially supported by the Simons Collaboration Grants on the Non-Perturbative Bootstrap and on Confinement and QCD Strings and by the DOE grant DE-SC0007870. DlZ is supported in part by the Royal Society University Research Fellowships grant URF/R1/221310 ``Bootstrapping Quantum Gravity''.

\appendix

\section{Conformal Transformation of a Curved Line}\label{sec:apd:confTransf}

In this section, we study conformal transformations of arbitrary paths. 
We parameterize the path by its projection on the $x^3$ axis,
\beq
\label{eq:genericpath}
x_a(s) = (\v_a^1(s), \v_a^2(s), s)\, ,
\eeq
assuming that the transverse vector function $\vec{\v}_a(s) \equiv \{\v_a^1(s), \v_a^2(s)\}$ is single valued and smooth. 
We then apply the following conformal transformation to it
\beq \label{eqn-apd-confGen}
x_a\quad\rightarrow\quad x_{ac} = \mathcal{R}_{\hat{x}^1}(\mathfrak{c}_1^2)\cdot\mathcal{R}_{\hat{x}^2}(\mathfrak{c}_1^1)\cdot\mathcal{B}_b(x_a) + \{\mathfrak{c}_0^1, \mathfrak{c}_0^2,a_3\}\,,
\eeq
where $\mathcal{R}_{\hat{n}}(\theta)$ denotes the finite rotation of angle $\theta$ around axis $\hat{n}$. $\mathcal{B}_b(x_a)$ denotes the special conformal transformation
\begin{equation}
  \mathcal{B}_b(x)^\mu \equiv \frac{x^\mu -b^\mu x^2}{1-2 b\cdot x + b^2 x^2}\,,\qquad {\rm with} \qquad b=(-\mathfrak{c}_2^1, -\mathfrak{c}_2^2 ,b_3)\,.
\end{equation}
The last term in (\ref{eqn-apd-confGen}) is a constant shift. The transverse vectors $\vec{\mathfrak{c}}_0$, $\vec{\mathfrak{c}}_1$, and $\vec{\mathfrak{c}}_2$ are a set of continuous parameters that we use to parameterize the conformal transformation \eqref{conftrans}. Here, we did not include rotations around the $\hat{x}^3$ axis, which leaves the straight line invariant. 

For generic values of the conformal transformation parameters, the two endpoints transform in all three directions. Using translation invariance, we can assume without loss of generality that the endpoint $x_0$ transforms only in the transverse direction. This choice fixes the value of $a_3$. We then further impose that the endpoint $x_1$ also transforms in the transverse direction. 
This condition fixes $b_3$ in terms of the other parameters. There are two solutions for $b_3$ and we pick the one that connects smoothly to the identity.\footnote{For $\v_a$ that does \textit{not} move the endpoints, $a_3=0$ and $b_3$ does not depend on the original path. Otherwise, $a_3$ and $b_3$ depend on $\v_a$ at the endpoints.}

After doing so, we are left with a conformal transformation that depends on three transverse vectors, $\vec{\mathfrak{c}}_1$, $\vec{\mathfrak{c}}_2$ and $\vec{\mathfrak{c}}_3$, just as we have in equation \eqref{conftrans}. We assume that these parameters are small and of the same order $\cO(\beta)$, and expand $x_{ac}$.

\paragraph{Leading Order.}

To leading order, the conformal transformation just added up to the $x_a$ path\footnote{To leading order there is no need to re-parameterize the deformed path.}
\beq \label{eqn-apd-conf-1stOrderV}
{x}_{ac} = x_a + ( \vec{\mathfrak{c}}_0 + \vec{\mathfrak{c}}_1 s + \vec{\mathfrak{c}}_2 s^2 , 0)+O(\beta^2)\, .
\eeq

\paragraph{Second Order.}

There is no second-order contribution to ${x}_{ac}$ for the conformal transformation we are considering\beq \label{eqn-apd-conf-1stOrderV2}
{x}_{ac} = x_a + ( \vec{\mathfrak{c}}_0 + \vec{\mathfrak{c}}_1 s + \vec{\mathfrak{c}}_2 s^2 , 0)+O(\beta^3)\, .
\eeq

\paragraph{Third Order.}

In general, the conformal transformation moves the points on the line in the transverse direction as well as along the line. As a result, at third order, we also have to perform a reparameterization of the path, keeping the deformation vector $\v_c$ transverse to the straight line. 

Another subtlety that first appears in the third order is that the deformation profile of the path due to the conformal transformation starts to depend on $\v_a$. It has the following decomposition 
\beq
\v_c=\v^{(1,0)}_c + \v^{(3,0)}_c+\v_c^{(1,2)} +  \v_c^{(2,1)} + \dots\,,
\eeq
where $\v_c^{(n,m)}\propto\beta^n\v_a^m$. The first-order term, $\v_c^{(1,0)}$, is given by the second term in \eqref{eqn-apd-conf-1stOrderV}. The third-order terms read
\beq
\begin{aligned}
    \left(\v^{(3,0)}_c\right)^\pm_s & = 2 {\mathfrak{c}_2^\mp} {\mathfrak{c}_2^\pm}^2 s^4+2 {\mathfrak{c}_2^\pm} s^3 ({\mathfrak{c}_2^\mp} {\mathfrak{c}_1^\pm}+{\mathfrak{c}_1^\mp} {\mathfrak{c}_2^\pm})\\
   & +\frac{1}{4} s^2 \left(2 {\mathfrak{c}_2^\mp} {\mathfrak{c}_1^\pm}^2+{\mathfrak{c}_2^\pm} \left(10 {\mathfrak{c}_1^\mp} {\mathfrak{c}_1^\pm}-{\mathfrak{c}_1^\mp}^2+{\mathfrak{c}_1^\pm}^2\right)\right)\\
    &+\frac{1}{24} s \left(-3 {\mathfrak{c}_1^\mp}^2 {\mathfrak{c}_1^\pm}+15 {\mathfrak{c}_1^\mp} {\mathfrak{c}_1^\pm}^2+{\mathfrak{c}_1^\mp}^3+3 {\mathfrak{c}_1^\pm}^3\right)\, ,\\
    \left(\v^{(1,2)}_c\right)^\pm_s & = s^2 {\dot{v}^\pm_0} ({\mathfrak{c}_1^\mp} {v^\pm_0}-({\mathfrak{c}_1^\mp}+2 {\mathfrak{c}_2^\mp}) {v^\pm_1}+{\mathfrak{c}_1^\pm} {v^\mp_0}-({\mathfrak{c}_1^\pm}+2 {\mathfrak{c}_2^\pm}) {v^\mp_1}) \\
    & +2 s \left({v^\pm_s} \left({\mathfrak{c}_2^\mp} {\dot{v}^\pm_0}-{\mathfrak{c}_1^\mp} {v^\pm_0}+({\mathfrak{c}_1^\mp}+2 {\mathfrak{c}_2^\mp}) {v^\pm_1}-{\mathfrak{c}_1^\pm} {v^\mp_0}+({\mathfrak{c}_1^\pm}+2 {\mathfrak{c}_2^\pm}) {v^\mp_1}\right)+{\mathfrak{c}_2^\pm} {v^\mp_s} {\dot{v}^\pm_0}\right)\\
    &+{\dot{v}^\pm_0} ({\mathfrak{c}_1^\mp} {v^\pm_s}-{\mathfrak{c}_1^\mp} {v^\pm_0}+{\mathfrak{c}_1^\pm} {v^\mp_s}-{\mathfrak{c}_1^\pm} {v^\mp_0})-2 {\mathfrak{c}_2^\mp} {v^\pm_s}^2\, , \\
    \left(\v^{(2,1)}_c\right)^\pm_s & = 2 {\mathfrak{c}_2^\mp} {\mathfrak{c}_2^\pm} s^3 {\dot{v}^\pm_0}+s^2 \left(2 {\mathfrak{c}_2^\pm}^2 {v^\mp_s}-({\mathfrak{c}_1^\mp} {\mathfrak{c}_1^\pm}+2 {\mathfrak{c}_2^\mp} {\mathfrak{c}_2^\pm}) {\dot{v}^\pm_0}\right) \\
    &+s \left({\mathfrak{c}_1^\mp} {\mathfrak{c}_1^\pm} {\dot{v}^\pm_0}+2 ({\mathfrak{c}_1^\mp}+{\mathfrak{c}_2^\mp}) ({\mathfrak{c}_1^\pm}+2 {\mathfrak{c}_2^\pm}) {v^\pm_s}-2 {\mathfrak{c}_2^\pm} ({\mathfrak{c}_1^\mp} {v^\pm_0}+{\mathfrak{c}_1^\pm} {v^\mp_0})+2 {\mathfrak{c}_1^\pm} {\mathfrak{c}_2^\pm} {v^\mp_s}\right)\\
    &+\frac{1}{4} \left(2 {\mathfrak{c}_1^\pm} ({\mathfrak{c}_1^\pm} ({v^\mp_s}-2 {v^\mp_0})-2 {\mathfrak{c}_1^\mp} {v^\pm_0})+\left(2 {\mathfrak{c}_1^\mp} {\mathfrak{c}_1^\pm}-{\mathfrak{c}_1^\mp}^2+{\mathfrak{c}_1^\pm}^2\right) {v^\pm_s}\right)\, .
\end{aligned}
\eeq
Here, for notational simplicity, we have omitted the index $a$ from $v_a$. In this context, the lower $s$-index of $v$ indicates the position. For example, $v_s^\pm \equiv v_a^\pm(s)$, and so on.

\section{Scheme-Independence of the Second Order Variation}
\label{apd:sec:2ndOrderSchemeInd}

The regularization scheme that we have used to regularize the double integral in (\ref{2-int}) is not unique. We can instead choose to perform the $s$ integral first,
\beq \label{eqn-apd-LeftReg}
\[\iint \mathcal{P}  \dd s\, \dd t\ldots\]_\text{reg 2} ={\ii \over2\sin(2\pi\Delta)} \ \supset\hspace{-5mm}\int\limits_{1,0)}\dd t \oint\limits_{[1,t]} \dd s\ldots\, .
\eeq
or to average the both orders,
\beq \label{eqn-apd-SymReg}
\[\iint \mathcal{P} \dd s\, \dd t\ldots\]_\text{reg 3} = \frac{1}{2}\times {\ii \over2\sin(2\pi\Delta)}  \Big[\supset\hspace{-5mm}\int\limits_{1,0)}\dd t \oint\limits_{[1,t]} \dd s\ldots +\subset\hspace{-5mm}\int\limits_{(1,0}\dd s \oint\limits_{[s,0]} \dd t\ldots \Big]\, .
\eeq

For the first regularization scheme \eqref{eqn-apd-LeftReg}, we find that the values of the invariants $\Lambda$, $\Xi$, and $\Sigma$ remain the same, see \eqref{LambdaXiSigma}. The values of the boundary terms, however, are swapped and can be obtained by exchanging $\gamma_j \leftrightarrow \tilde{\gamma}_j$ in \eqref{eqn-2ndOrder-AllValueGamma}.

For the second symmetric regularization scheme \eqref{eqn-apd-SymReg} we can repeat the bootstrap above. The values of the invariants $\Lambda$, $\Xi$, and $\Sigma$ remain the same as in \eqref{LambdaXiSigma}. This scheme preserves the CPT symmetry, which interchanges the left-right ends. Thus, the resulting boundary terms are symmetric. They are given by
\beq
\begin{aligned}
    \gamma_0 & = \tilde \gamma_0 = - \frac{1}{4}\,, \qquad & \gamma_1 & = \tilde \gamma_1 = \frac{3-2\Delta}{8}\,, \\
    \gamma_2 & = \tilde \gamma_2 = \frac{\Delta(3-2\Delta)}{8}, \qquad & \gamma_3 & = \tilde \gamma_3 = 0, \\
    \gamma_4 & = \tilde \gamma_4 =\frac{1+2\Delta}{8}\,, \qquad & \gamma_5 & = \tilde \gamma_5 = \frac{1}{8}(2\Delta^2 -3\Delta-2)\, . \\
\end{aligned}
\eeq

From the two examples above, we can see that different regularization schemes of the double integral result in different values for the boundary coefficients.

\bibliography{bib}

\providecommand{\href}[2]{#2}\begingroup\raggedright\begin{thebibliography}{10}

\bibitem{Affleck:1991tk}
I.~Affleck and A.~W.~W. Ludwig, {\it {Universal noninteger 'ground state degeneracy' in critical quantum systems}},  {\em Phys. Rev. Lett.} {\bf 67} (1991) 161--164.

\bibitem{Polchinski:2011im}
J.~Polchinski and J.~Sully, {\it {Wilson Loop Renormalization Group Flows}},  {\em JHEP} {\bf 10} (2011) 059, [\href{http://arxiv.org/abs/1104.5077}{{\tt arXiv:1104.5077}}].

\bibitem{Friedan:2003yc}
D.~Friedan and A.~Konechny, {\it {On the boundary entropy of one-dimensional quantum systems at low temperature}},  {\em Phys. Rev. Lett.} {\bf 93} (2004) 030402, [\href{http://arxiv.org/abs/hep-th/0312197}{{\tt hep-th/0312197}}].

\bibitem{Casini:2016fgb}
H.~Casini, I.~Salazar~Landea, and G.~Torroba, {\it {The g-theorem and quantum information theory}},  {\em JHEP} {\bf 10} (2016) 140, [\href{http://arxiv.org/abs/1607.00390}{{\tt arXiv:1607.00390}}].

\bibitem{Cuomo:2021rkm}
G.~Cuomo, Z.~Komargodski, and A.~Raviv-Moshe, {\it {Renormalization Group Flows on Line Defects}},  {\em Phys. Rev. Lett.} {\bf 128} (2022), no.~2 021603, [\href{http://arxiv.org/abs/2108.01117}{{\tt arXiv:2108.01117}}].

\bibitem{Kazakov:1980zj}
V.~A. Kazakov, {\it {WILSON LOOP AVERAGE FOR AN ARBITRARY CONTOUR IN TWO-DIMENSIONAL U(N) GAUGE THEORY}},  {\em Nucl. Phys. B} {\bf 179} (1981) 283--293.

\bibitem{Giombi:2009ms}
S.~Giombi, V.~Pestun, and R.~Ricci, {\it {Notes on supersymmetric Wilson loops on a two-sphere}},  {\em JHEP} {\bf 07} (2010) 088, [\href{http://arxiv.org/abs/0905.0665}{{\tt arXiv:0905.0665}}].

\bibitem{Drukker:1999zq}
N.~Drukker, D.~J. Gross, and H.~Ooguri, {\it {Wilson loops and minimal surfaces}},  {\em Phys. Rev. D} {\bf 60} (1999) 125006, [\href{http://arxiv.org/abs/hep-th/9904191}{{\tt hep-th/9904191}}].

\bibitem{Polyakov:2000jg}
A.~M. Polyakov and V.~S. Rychkov, {\it {Loop dynamics and AdS / CFT correspondence}},  {\em Nucl. Phys. B} {\bf 594} (2001) 272--286, [\href{http://arxiv.org/abs/hep-th/0005173}{{\tt hep-th/0005173}}].

\bibitem{Polyakov:2000ti}
A.~M. Polyakov and V.~S. Rychkov, {\it {Gauge field strings duality and the loop equation}},  {\em Nucl. Phys. B} {\bf 581} (2000) 116--134, [\href{http://arxiv.org/abs/hep-th/0002106}{{\tt hep-th/0002106}}].

\bibitem{Semenoff:2004qr}
G.~W. Semenoff and D.~Young, {\it {Wavy Wilson line and AdS / CFT}},  {\em Int. J. Mod. Phys. A} {\bf 20} (2005) 2833--2846, [\href{http://arxiv.org/abs/hep-th/0405288}{{\tt hep-th/0405288}}].

\bibitem{Giombi:2009ek}
S.~Giombi and V.~Pestun, {\it {The 1/2 BPS 't Hooft loops in N=4 SYM as instantons in 2d Yang-Mills}},  {\em J. Phys. A} {\bf 46} (2013) 095402, [\href{http://arxiv.org/abs/0909.4272}{{\tt arXiv:0909.4272}}].

\bibitem{Toledo:2014koa}
J.~C. Toledo, {\it {Smooth Wilson loops from the continuum limit of null polygons}},  \href{http://arxiv.org/abs/1410.5896}{{\tt arXiv:1410.5896}}.

\bibitem{Cooke:2017qgm}
M.~Cooke, A.~Dekel, and N.~Drukker, {\it {The Wilson loop CFT: Insertion dimensions and structure constants from wavy lines}},  {\em J. Phys. A} {\bf 50} (2017), no.~33 335401, [\href{http://arxiv.org/abs/1703.03812}{{\tt arXiv:1703.03812}}].

\bibitem{Gabai:2022vri}
B.~Gabai, A.~Sever, and D.-l. Zhong, {\it {Line Operators in Chern-Simons\textendash{}Matter Theories and Bosonization in Three Dimensions}},  {\em Phys. Rev. Lett.} {\bf 129} (2022), no.~12 121604, [\href{http://arxiv.org/abs/2204.05262}{{\tt arXiv:2204.05262}}].

\bibitem{Gabai:2022mya}
B.~Gabai, A.~Sever, and D.-l. Zhong, {\it {Line operators in Chern-Simons-Matter theories and Bosonization in Three Dimensions II: Perturbative analysis and all-loop resummation}},  {\em JHEP} {\bf 04} (2023) 070, [\href{http://arxiv.org/abs/2212.02518}{{\tt arXiv:2212.02518}}].

\bibitem{Cardy:1996xt}
J.~L. Cardy, {\em {Scaling and renormalization in statistical physics}}.
\newblock 1996.

\bibitem{Zamolodchikov:1987ti}
A.~B. Zamolodchikov, {\it {Renormalization Group and Perturbation Theory Near Fixed Points in Two-Dimensional Field Theory}},  {\em Sov. J. Nucl. Phys.} {\bf 46} (1987) 1090.

\bibitem{Kutasov:1988xb}
D.~Kutasov, {\it {Geometry on the Space of Conformal Field Theories and Contact Terms}},  {\em Phys. Lett. B} {\bf 220} (1989) 153--158.

\bibitem{Ranganathan:1993vj}
K.~Ranganathan, H.~Sonoda, and B.~Zwiebach, {\it {Connections on the state space over conformal field theories}},  {\em Nucl. Phys. B} {\bf 414} (1994) 405--460, [\href{http://arxiv.org/abs/hep-th/9304053}{{\tt hep-th/9304053}}].

\bibitem{Gaberdiel:2008fn}
M.~R. Gaberdiel, A.~Konechny, and C.~Schmidt-Colinet, {\it {Conformal perturbation theory beyond the leading order}},  {\em J. Phys. A} {\bf 42} (2009) 105402, [\href{http://arxiv.org/abs/0811.3149}{{\tt arXiv:0811.3149}}].

\bibitem{Amoretti:2017aze}
A.~Amoretti and N.~Magnoli, {\it {Conformal perturbation theory}},  {\em Phys. Rev. D} {\bf 96} (2017), no.~4 045016, [\href{http://arxiv.org/abs/1705.03502}{{\tt arXiv:1705.03502}}].

\bibitem{Behan:2017mwi}
C.~Behan, {\it {Conformal manifolds: ODEs from OPEs}},  {\em JHEP} {\bf 03} (2018) 127, [\href{http://arxiv.org/abs/1709.03967}{{\tt arXiv:1709.03967}}].

\bibitem{Cavaglia:2022yvv}
A.~Cavagli\`a, N.~Gromov, J.~Julius, and M.~Preti, {\it {Integrated correlators from integrability: Maldacena-Wilson line in $ \mathcal{N} $ = 4 SYM}},  {\em JHEP} {\bf 04} (2023) 026, [\href{http://arxiv.org/abs/2211.03203}{{\tt arXiv:2211.03203}}].

\bibitem{Hartman:2022zik}
T.~Hartman, D.~Mazac, D.~Simmons-Duffin, and A.~Zhiboedov, {\it {Snowmass White Paper: The Analytic Conformal Bootstrap}},  in {\em {Snowmass 2021}}, 2, 2022.
\newblock \href{http://arxiv.org/abs/2202.11012}{{\tt arXiv:2202.11012}}.

\bibitem{Poland:2022qrs}
D.~Poland and D.~Simmons-Duffin, {\it {Snowmass White Paper: The Numerical Conformal Bootstrap}},  in {\em {Snowmass 2021}}, 3, 2022.
\newblock \href{http://arxiv.org/abs/2203.08117}{{\tt arXiv:2203.08117}}.

\bibitem{Witten:1988hf}
E.~Witten, {\it {Quantum Field Theory and the Jones Polynomial}},  {\em Commun. Math. Phys.} {\bf 121} (1989) 351--399.

\bibitem{Vasiliev:1992av}
M.~A. Vasiliev, {\it {More on equations of motion for interacting massless fields of all spins in (3+1)-dimensions}},  {\em Phys. Lett. B} {\bf 285} (1992) 225--234.

\bibitem{Giombi:2011kc}
S.~Giombi, S.~Minwalla, S.~Prakash, S.~P. Trivedi, S.~R. Wadia, and X.~Yin, {\it {Chern-Simons Theory with Vector Fermion Matter}},  {\em Eur. Phys. J. C} {\bf 72} (2012) 2112, [\href{http://arxiv.org/abs/1110.4386}{{\tt arXiv:1110.4386}}].

\bibitem{Aharony:2011jz}
O.~Aharony, G.~Gur-Ari, and R.~Yacoby, {\it {d=3 Bosonic Vector Models Coupled to Chern-Simons Gauge Theories}},  {\em JHEP} {\bf 03} (2012) 037, [\href{http://arxiv.org/abs/1110.4382}{{\tt arXiv:1110.4382}}].

\bibitem{Aharony:2012nh}
O.~Aharony, G.~Gur-Ari, and R.~Yacoby, {\it {Correlation Functions of Large N Chern-Simons-Matter Theories and Bosonization in Three Dimensions}},  {\em JHEP} {\bf 12} (2012) 028, [\href{http://arxiv.org/abs/1207.4593}{{\tt arXiv:1207.4593}}].

\bibitem{Klebanov:2002ja}
I.~R. Klebanov and A.~M. Polyakov, {\it {AdS dual of the critical O(N) vector model}},  {\em Phys. Lett. B} {\bf 550} (2002) 213--219, [\href{http://arxiv.org/abs/hep-th/0210114}{{\tt hep-th/0210114}}].

\bibitem{Sezgin:2003pt}
E.~Sezgin and P.~Sundell, {\it {Holography in 4D (super) higher spin theories and a test via cubic scalar couplings}},  {\em JHEP} {\bf 07} (2005) 044, [\href{http://arxiv.org/abs/hep-th/0305040}{{\tt hep-th/0305040}}].

\bibitem{Chang:2012kt}
C.-M. Chang, S.~Minwalla, T.~Sharma, and X.~Yin, {\it {ABJ Triality: from Higher Spin Fields to Strings}},  {\em J. Phys. A} {\bf 46} (2013) 214009, [\href{http://arxiv.org/abs/1207.4485}{{\tt arXiv:1207.4485}}].

\bibitem{Leigh:2003gk}
R.~G. Leigh and A.~C. Petkou, {\it {Holography of the N=1 higher spin theory on AdS(4)}},  {\em JHEP} {\bf 06} (2003) 011, [\href{http://arxiv.org/abs/hep-th/0304217}{{\tt hep-th/0304217}}].

\end{thebibliography}\endgroup
\bibliographystyle{JHEP}

\end{document}